\documentstyle[prd,aps,epsf]{revtex}

\begin{document}

\title{Scattering of scalar waves by rotating black holes}

\author{Kostas Glampedakis$^{1}$ and Nils Andersson$^{2}$}

\address{$^1$ Department of Physics and Astronomy,
 Cardiff University, Cardiff CF2 3YB, United Kingdom}
\address{$^2$ Department of Mathematics, 
University of Southampton, Southampton SO17 1BJ, United Kingdom}

\maketitle

\begin{abstract}

We study the scattering of massless scalar waves by a Kerr black hole
by letting  plane monochromatic waves impinge on the black hole.
We calculate the relevant scattering phase-shifts using the
 Pr\"{u}fer phase-function method, which is
computationally efficient and reliable also for high 
frequencies and/or large values of the angular multipole indices ($l$,$m$).
We use the obtained phase-shifts and the
partial-wave approach to determine  differential cross sections 
and deflection functions. Results for  off-axis scattering (waves 
incident along directions misaligned with the black hole's rotation
axis)  are obtained for the first time. 
Inspection of the off-axis deflection functions reveals the same scattering
phenomena as in Schwarzschild scattering.
In particular, the cross sections are dominated by the glory effect
and the forward (Coulomb) divergence due to the long-range nature of 
the gravitational field. In the rotating case
the overall diffraction
pattern is ``frame-dragged'' and as a result the glory maximum is 
not observed in the exact backward direction. We discuss the physical reason
for this behaviour, and explain it in terms of the distinction between 
prograde and retrograde motion in the Kerr 
gravitational field. Finally, we also discuss the possible influence of
the so-called superradiance effect on the scattered waves. 
\end{abstract}


\section{INTRODUCTION}

Diffraction of scattered waves provides the explanation for 
many of Nature's most beautiful phenomena, 
such as rainbows and glories. It has long been recognized that 
these optical phenomena have analogies in many other branches of physics. 
They are of particular relevance to quantum physics, where
plane wave ``beams'' are routinely used to probe the details of  
atoms, nuclei  or molecules. Such experiments provide a 
deep understanding of the scatterer's physics and can be used as a 
powerful test of various theoretical models. The 
analogy can be extended also to gravitational physics and extreme
astrophysical objects like black holes. In fact, black hole scattering 
has been the subject of a considerable amount
of work carried out over the last 30 years 
(see \cite{book} for an extensive review). In the case of
astrophysical black holes it is unlikely that the various 
diffraction effects will ever be observed (although it is not
entirely implausible that advances of current technology will
eventually enable us to study interference effects in  
gravitationally lensed waves). However, it is nevertheless 
useful to have a detailed theoretical understanding of
the scattering of waves from black holes.
After all, a study of these problems provides a deeper insight 
into the physics of black holes as well as wave-propagation 
in curved spacetimes. 

The benchmark problem for black-hole scattering is 
massless scalar waves impinging on a Schwarzshild black hole.
This problem is well understood 
\cite{Ryan,Sanchez1,Sanchez2,Sanchez3,Sanchez4,path_int1,Nils}, and
it is known that it provides a beautiful example of the glory effect.
Handler and Matzner \cite{Handler} have shown that
the situation remains almost unchanged if, instead of scalar waves,
 one decides to ``shoot'' plane electromagnetic or 
gravitational waves towards the black hole. These authors have also
considered on-axis scattering of gravitational waves
in the case when the black hole is rotating \cite{Handler}. 
Their results suggest that the scattering cross sections consist of
essentially
the same features as in the non-rotating case (there is a
forward divergence due to the long-range nature of the gravitational field
and a backward glory). In addition, they find some peculiar 
features that are, at the present time, not well
understood. An explanation of these effects is complicated by the fact that 
they could be caused by several effects, the most important
being the 
coupling between the black hole's spin and the 
spin/polarisation of the incident wave. Given that the available
investigations have not been able to distinguish between these
various effects, we feel that our current understanding 
is somewhat unsatisfactory. This feeling is enhanced by the 
fact that no results for the most realistic case, corresponding 
to off-axis incidence, have yet been obtained.

This paper provides an attempt to further our understanding 
of the scattering from rotating black holes. Our aim is to isolate 
those scattering effects that are due to the spin of the black hole. 
In order to do this, we focus our attention on the  
scattering of massless scalar waves. For this case, the infalling waves
have neither spin nor polarisation and therefore one would expect the
scattered wave to have a simpler character than in the physically
more relevant case of gravitational waves. However, one can be 
quite certain that the features discussed in this paper will be present 
also in the case of gravitational waves. It is, after all, well known that 
the propagation of various fields in a given black hole geometry 
is described by very similar wave equations.

Although we will re-examine the case of axially incident waves,  
our main attention will be on the more interesting off-axis
scattering cross sections. These cross sections turn out to 
be quite different from the ones available in the 
literature. Obviously, they have two 
degrees of freedom (corresponding to
the two angles $\theta$ and $\varphi$ in  
Boyer-Lindquist coordinates). In addition we will show that the 
cross sections are  asymmetric with respect to the incidence direction. 
In particular, the glory moves away from the backward direction as 
a result of rotational frame dragging that provides a 
distinction between prograde and retrograde motion in the Kerr geometry.

We construct our differential cross sections using the 
well-known partial wave decomposition (for an introduction see 
\cite{Newton}) --- the standard approach in quantum scattering theory. 
That this method is equally useful in black-hole scattering 
is well established \cite{book}. We should point out, however, 
that alternatives (such as the complex-angular momentum approach 
\cite{CAM1,CAM2} 
and path-integral methods \cite{path_int1,path_int2,path_int3}) 
have also been succesfully applied to the black-hole case.
In the partial wave picture all scattering   information  is 
contained in the radial wavefunction's phase shifts. The calculation of these
phase-shifts must, apart from in  
exceptional cases like Coulomb scattering,
be performed numerically. Various techniques have been developed for this 
task. Basically, one must be able to determine the phase-shifts accurately
up to sufficiently large $l$ partial waves that no 
interference effects are lost. This boils down to a need for 
many more multipoles to be studied as the frequency of the infalling wave is 
increased. In black-hole scattering several methods have been employed
for the phase-shift calculation:
Matzner and Ryan \cite{Ryan} numerically integrated the 
relevant radial wave equation (Teukolsky's equation). 
Since the desired solution is an 
oscillating function, this calculation becomes increasingly difficult (and 
time consuming) as the frequency is increased. Consequently, Matzner and 
Ryan restricted their study of electromagnetic and gravitational wave
scattering to $\omega M \leq 0.75 $ and $ l \leq 10 $. In order to avoid 
this difficulty, Handler and Matzner \cite{Handler} combined a 
numerical solution in the region where the gravitational curvature 
potential varies rapidly, with an approximate WKB solution for 
relatively large values of the radial coordinate. This trick allowed 
them to perform calculations for $l \leq 20 $ and $\omega M \leq 2.5 $.
Some years ago one of us  used the phase-integral
method \cite{Froman1,Froman2} to derive an approximate formula 
for the phase-shifts in the context of Schwarzschild scattering \cite{Nils}.
This formula was shown to be reliable and efficient even for high 
frequencies and/or large $l$ values (in \cite{Nils}
results for $\omega M= 10 $ and $l \leq 200 $ were presented).
This means that the differential cross sections determined from the
phase-integral phase-shifts were reliable also for rather high frequencies. 
Even though the phase-integral formula could be generalised to 
scattering by a Kerr black hole and therefore used for the
purposes of the present study,
we have chosen a different approach here. Our phase-shift
determination is based on the 
so-called Pr\"ufer method (well-known in quantum scattering 
theory \cite{Pajunen1,Pajunen2} and, in general, in numerical treatments 
of Sturm-Liouville problems \cite{Pryce} ) which, in a nutshell, 
involves transforming the original radial 
wavefunction to specific phase-functions
and numerical integration of the resulting equations. In essence, this
method is a close relative of the phase-amplitude method that was devised
by one of us to study black-hole resonances \cite{pam}.

The remainder of the paper is organised as follows. 
In Sections IIA and IIB the problem of scattering  by a Kerr black hole is 
rigorously formulated. In Section IIC 
the important notion of the deflection function is discussed. 
Section III is devoted to our numerical results.  
First, in Section IIIA our numerical method for calculating phase-shifts 
is presented. In Section IIIB familiar Schwarzschild results are 
reproduced as a code validation.  Sections IIIC and IIID contain 
entirely new information: Differential cross sections and 
deflection functions for on and off-axis scattering respectively. 
These are the main results of the paper. Furthermore, in Section 
IIIE we present
numerical results concerning forward glories.
 The role of superradiance for scattering of monochromatic waves 
is discussed in Section IIIF. Our conclusions are briefly
summarised in Section IV. Three appendices are devoted to
technical details, which are included for completeness.
In Appendix A we discuss the notion of 
``plane waves''  in the presence of a gravitational field. 
In Appendix B the partial-wave decomposition of a plane wave 
in the Kerr background is determined, and finally in Appendix C 
we briefly describe the method we have used to calculate the spin-$0$ 
spheroidal harmonics and their eigenvalues. Throughout the paper 
we adopt geometrised units ($c=G=1$).


\section{Scattering from black holes}

\subsection{Formulation of the problem}

We consider a massless scalar field in the  Kerr black-hole geometry. 
Then, first-order black-hole perturbation theory, basically the Teukolsky 
equation \cite{Teukolsky}, applies. The scalar field satisfies the curved
spacetime wave equation $ \Box \Phi =0$. Adopting standard Boyer-Lindquist
coordinates we can always decompose the field
as (since the spacetime is axially symmetric)
 \begin{equation}
\Phi(r,\theta,\varphi,t)= \frac{1}{\sqrt{r^2 +a^2}}\sum_{m=-\infty}^{+\infty}
\phi_{m}(r, \theta,t) e^{im\varphi}
\label{eq1}
\end{equation}
In scattering problems it is customary to consider monocromatic waves
with given frequency $\omega$. Therefore we can further write
\begin{equation}
\phi_{m}(r,\theta,t)= \sum_{l=|m|}^{+\infty} c_{lm} u_{lm}(r,\omega) S_{lm}^{a\omega}
(\theta) e^{-i\omega t}
\label{eq2}
\end{equation}
where $c_{lm}$ is some expansion coefficient and $S_{lm}^{a\omega}(\theta)$ 
are the usual spin-0 spheroidal harmonics. These are normalised as
\begin{equation}
\int_{0}^{\pi} d\theta \sin\theta |S_{lm}^{a\omega}(\theta)|^2 = 
\frac{1}{2\pi}
\end{equation}
Finally, the  function $u_{lm}(r,\omega) $ is a 
solution of the radial Teukolsky
equation:
\begin{equation}
{d^2 u_{lm} \over dr_\ast^2} + \left[ {K^2  + (2am\omega
-a^2\omega^2 - E_{lm})\Delta \over (r^2+a^2)^2} -{dG \over dr_\ast} -
G^2 \right] u_{lm} = 0 
\label{Teuk}
\end{equation} 
where $K=(r^2 + a^2)\omega - am$ and $G=r\Delta/(r^2+a^2)^2$. Furthermore,
 $E_{lm}$ denotes the angular eigenvalue, cf. Appendix~C. 
As usual, $\Delta= r^2 -2Mr + a^2$ and 
the ``tortoise'' radial coordinate $ r_{\ast}$
is defined as (with $r_{\pm}$, the two solutions to $\Delta =0$, 
denoting the event horizon and the inner Cauchy 
horizon of the black hole)
\begin{equation}
r_{\ast} = r + \frac{2Mr_{+}}{r_{+}-r_{-}}\ln \left ( \frac{r}{r_{+}} -1
\right ) -\frac{2Mr_{-}}{r_{+}-r_{-}}\ln \left ( \frac{r}{r_{-}} -1
\right ) + c
\label{rstar}
\end{equation}
Usually, the arbitrary integration constant $c$ is disregarded in this
relation. However, in scattering problems it turns out to be useful
to keep it, as we shall see later.

We are interested in a causal solution to (\ref{Teuk})
which describes waves that are purely ``ingoing''
at the black hole's horizon. This solution can be written
\begin{equation}
u_{ lm}^{\rm in} \sim \left\{ \begin{array} {ll}
e^{-ikr_\ast} \quad \mbox{as } r\to r_+ \ , \\ A^{\rm out}_{lm} 
e^{i\omega r_\ast} + A^{\rm in}_{lm} e^{-i\omega r_\ast} \quad
\mbox{as } r\to +\infty \ .
\end{array} \right.
\label{uin}\end{equation}
where $k= \omega -ma/2Mr_{+}= \omega - m\omega_+$.
In addition, we want to  impose an ``asymptotic scattering
boundary condition''. We want the total field at spatial infinity 
to be the
sum of a plane wave plus an outgoing scattered wave. In other words, 
we should have 
\begin{equation}
\Phi(r,\theta,\varphi) \sim \Phi_{\rm plane} + \frac{1}{r}
f(\theta,\varphi) e^{i\omega r_{\ast}} \quad \mbox{as} \quad r \to +\infty
\label{condition}  
\end{equation}
where we have omitted the trivial time-dependence.
All information regarding scattering is contained in the 
(complex-valued) scattering amplitude $f(\theta,\varphi)$. 
Note that, unlike in axially symmetric
scattering the scattering amplitude will depend on both angles: $\theta$ and
$\varphi$. 

Up to this point, we have used the term ``plane wave'' quite loosely.
In the presence of a long-range field such as the Kerr gravitational field
(which falls off as $\sim 1/r $ at infinity) we cannot write a plane
wave in the familiar flat space form.  This problem has been discussed
in several papers, see \cite{Matzner,Chrzanowski}. 
Remarkably, it turns out that in a black hole background 
 the long-range character of the field is accounted 
for by a logarithmic phase-modification of the
flat space plane-wave expression. In practice, the substitution 
$ r \to r_{\ast} $ is made in the various exponentials.
 In order to make this paper as self-contained as possible, 
we discuss this point in some detail in Appendix~A.

\begin{figure}[tbh]
\centerline{ \epsfysize=5cm \epsfbox{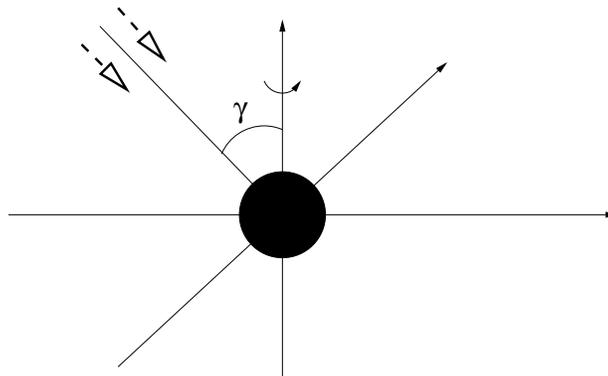}}
\caption{A schematic illustration of the general scattering problem. A plane
wave impinges on a rotating black hole making an angle $\gamma$ 
with the rotation axis.}
\label{draw1}\end{figure}

The asymptotic expression for a plane wave travelling along a 
direction making an angle $\gamma $ with the black hole's spin axis, see 
Figure~\ref{draw1}, is  
\begin{equation}
\Phi_{\rm plane}= e^{i\omega r_{\ast}(\sin\gamma \sin\theta \sin\varphi +
 \cos\gamma
\cos\theta)}
\label{ofplane}
\end{equation}
where without any loss of generality we have assumed an amplitude of
unity. We can decompose this plane wave 
in a way similar to (\ref{eq1}) and (\ref{eq2});
\begin{equation}
\Phi_{\rm plane} \approx \frac{1}{r}\sum_{m=-\infty}^{+\infty}
\sum_{l=|m|}^{+\infty} c_{lm}^{(0)} u_{lm}^{(0)}(r,\omega) 
S_{lm}^{a\omega}(\theta)e^{im\varphi}
\end{equation}
where $u_{lm}^{(0)}$ are asymptotic solutions of (\ref{Teuk}).
For $r \to \infty$ we have (see Appendix B),
\begin{equation}
c_{lm}^{(0)}u_{lm}^{(0)}(r,\omega) \approx 2\pi S_{lm}^{a\omega}(\gamma)
\left \{ (-i)^{m+1} e^{i\omega r_{\ast}} + i^{m+1} (-1)^{l+m} 
e^{-i\omega r_{\ast}} \right \}
\label{plane_asym}
\end{equation}
Similarly, the full field at infinity can be approximated as :
\begin{equation}
\Phi \approx \frac{1}{r}\sum_{m=-\infty}^{+\infty}\sum_{l=|m|}^{+\infty}
c_{lm}(A^{\rm in}_{lm} e^{-i\omega r_{\ast}}+ A^{\rm out}_{lm}e^{i\omega r_{\ast}})S_{lm}^{a\omega}(\theta) e^{im\varphi}
\end{equation}
By imposing the scattering condition (\ref{condition}), we can fix $c_{lm}$
by demanding that the ingoing wave piece of $\Phi -\Phi_{\rm plane}$  
vanishes. 
After some straightforward manipulations we get for the scattering amplitude
\begin{equation}
f(\theta,\varphi)= \frac{2\pi}{i\omega}\sum_{m=-\infty}^{+\infty}
\sum_{l=|m|}^{+\infty}(-i)^{m}S_{lm}^{a\omega}(\theta)S_{lm}^{a\omega}(\gamma)
e^{im\varphi}\left [ (-1)^{l+1} \frac{A^{\rm out}_{lm}}{A^{\rm in}_{lm}} -1 
\right ]
\label{amp_off}
\end{equation}
By defining  the ``scattering matrix element'' 
${\cal S}_{lm}=(-1)^{l+1} A^{\rm out}_{lm}/A^{\rm in}_{lm}$ we can
equivalently write
\begin{equation}
{\cal S}_{lm}= e^{2i\delta_{lm}}
\end{equation}
where we have introduced the phase-shift $\delta_{lm}$. 
Thus we see that the phase-shifts $\delta_{lm}$ contain all relevant
information regarding the scattered wave. It is worth emphasising
that for non-axisymmetric scattering the phase-shifts will
depend on both $l$ and $m$. Also, the 
 $\delta_{lm}$ are in general complex valued in order
to account for absorption by the black hole. 
For later convenience, we also point out that the full field at infinity 
can be written
\begin{equation}
\Phi \sim \sin(\omega r_{\ast} + \delta_{lm} -\frac{l\pi}{2})
\quad \mbox{ as } r_\ast \to \infty
\end{equation}

In the case of on-axis incidence  ($\gamma=0$) the scattering 
amplitude simplifies considerably, and we get
\begin{equation}
f(\theta)= \frac{2\pi}{i\omega}\sum_{l=0}^{+\infty}
S_{l0}^{a\omega}(\theta)S_{l0}^{a\omega}(0)
\left [ (-1)^{l+1} \frac{A^{\rm out}_{l}}{A^{\rm in}_{l}} -1 \right ]
\label{amp_on} 
\end{equation}
Here we see that the outcome is no longer dependent on $m$, which is 
natural given the axial symmetry of the problem. Furthermore, it is 
easy to see that we  recover the familiar Schwarzschild expression \cite{Nils}
by setting $a=0$.

The differential cross section (often simply called the cross section 
in this paper) is the most important ``observable'' in a scattering problem.
It provides a  measure of the extent to which  the scattering target is 
``visible''  from a certain viewing angle. 
As demonstrated in standard textbooks
\cite{Newton}, the differential cross section follows immediately 
from the scattering amplitude
\begin{equation}
\frac{d\sigma}{d\Omega}= |f(\theta,\varphi)|^2 \ .
\label{dcs}
\end{equation}

This cross section corresponds to ``elastic'' scattering only, 
that is, it describes the angular distribution of the waves escaping 
to infinity. We
can similarly define an ``absorption cross section'' but we shall not be
concerned with this issue here. 
Nevertheless, as  we have already pointed out, 
black hole absorption has an effect on the phase-shifts that are
used to compute the cross section (\ref{dcs}).

The strategy then for a cross section calculation  (for given black 
hole parameters and wave frequency) involves three steps: i) calculation of the
phase-shifts $\delta_{lm}$, (or, equivalently, of the asymptotic
amplitudes $A_{lm}^{\rm out/in}$), ii) calculation of the spheroidal 
harmonics $S_{lm}^{a\omega}(\theta)$, and finally iii) evaluation   
of the sums in (\ref{amp_off}) and/or (\ref{amp_on}) 
including a sufficiently large number of terms.


\subsection{Approximating the scattering amplitude}

In practice, the partial-wave sum calculation is problematic as
it converges slowly.  In fact,  the sum is divergent for some angles. 
This is just an artifact due to the long-range nature of the 
gravitational field. No matter how far from the black hole a 
partial wave may travel, it will always ``feel'' the presence of the 
gravitational potential (that falls off as $1/r$). 
A similar behaviour is known to exist in Coulomb scattering.  
The divergence always occurs at the angle that
specifies the incident wave's propagation direction.
That this will be the case is easily  seen from the identity
\begin{equation}
\sum_{l,m} S_{lm}^{a\omega}(\theta)e^{im\varphi}S_{lm}^{a\omega}(\gamma)
e^{-im\pi/2}= \delta(\cos\theta -\cos\gamma)\delta(\varphi-\pi/2) \ ,
\label{orth}
\end{equation}
which follows directly from the fact that the functions 
$ S_{lm}^{a\omega}(\theta) e^{im\varphi}$ 
form an orthonormal set. The corresponding identity for on axis incidence is
\begin{equation}
\sum_{l} S_{l0}^{a\omega}(\theta)S_{l0}^{a\omega}(0)
= \frac{1}{2\pi}\delta(\cos\theta -1) \ .
\end{equation}
From  (\ref{orth}) we can deduce a peculiar feature:  Although the 
scattering problem is physically insensitive to the actual
$\varphi$ of the
incidence direction, the specific value $\varphi=\pi/2$ is imposed by the 
above relations. Of course, this has no physical relevance since the problem 
at hand is axially symmetric and we can, without any loss of generality,  
assume
an incoming wave travelling along the direction 
$(\theta,\varphi)=(\gamma,\pi/2)$.

The fact that the Kerr gravitational field behaves asymptotically as a 
Newtonian one considerably simplifies the scattering amplitude calculation.
We would expect that large $l$ partial waves (strictly speaking when
$l/\omega M \gg 1 $) to essentially feel only the far-zone Newtonian field.
In terms of the phase-shifts, we expect them to 
approach their Newtonian counterparts $ \delta_{lm} \to \delta_{l}^{N}$
asymptotically.
In order to secure this matching we add to our phase-shifts
 an ``integration constant'' $-2\omega M \ln(4\omega M) + \omega M $. 
In this way, we also get $ r_{\ast} \to r_{\rm c} $, where 
$r_{\rm c}= r + 2M\ln(2\omega r)$ is the respective tortoise coordinate of 
the Coulomb/Newtonian problem. 
Such a manipulation is admissible given the arbitrariness in the 
choice of the constant $c$ in (\ref{rstar}). 

In calculating the partial-wave sum for the scattering amplitude, 
it is convenient to split it into two terms:
\begin{equation}
f(\theta,\varphi)= f_D(\theta,\varphi) + f_N(\theta,\varphi)
\end{equation}  
Here, $f_D(\theta,\varphi)$ represents the part of the scattering amplitude
that carries the information of the main diffraction effects, while
$f_{N}(\theta,\varphi)$ denotes the Newtonian (Coulomb) amplitude.
Explicitly we have
\begin{equation}
f_N(\theta,\varphi)= \frac{2\pi}{i\omega} \sum_{l,m} Y_{lm}(\theta)
e^{im\varphi} Y_{lm}(\gamma) (-i)^m \left [ e^{2i\delta_l^N} -1 \right ]
\end{equation}
where we have deliberately ``forgotten'' the spherical symmetry of the
Newtonian potential (which would had allowed us to write $f_{N}$ as a 
function of $\theta$ only, and thus in terms of a sum over $l$).
However, the Newtonian phase-shifts $\delta_l^{N}$ are still given
by the well-known expression \cite{book},
\begin{equation}
e^{2i\delta_l^N}= \frac{\Gamma(l+1-2i\omega M)}
{\Gamma(l+1+2i\omega M)}
\end{equation} 
After simple manipulations we get
\begin{equation}
f_N(\xi)= \frac{1}{2i\omega}\sum_{l=0}^{+\infty} (2l+1)P_l(\cos\xi)
(e^{2i\delta_l^N} -1)
\label{fN2}
\end{equation}
where $\cos\xi= \cos\theta\cos\gamma + \sin\theta \sin\gamma \sin\varphi $. 
The sum in (\ref{fN2}) is known in closed form \cite{book};
\begin{equation}
f_N(\xi)= M\frac{\Gamma(1-2i\omega M)}{\Gamma(1+2i\omega M)}
\left [\sin\frac{\xi}{2} \right ]^{-2+4i\omega M}
\label{fN}
\end{equation}
From this we can see that $f_N(\theta,\varphi)$ diverges 
in the $\xi=0$ direction.

Let us now focus on the ``diffraction'' amplitude $f_{D}(\theta,\varphi)$.
It has the form
\begin{equation}
f_D(\theta,\varphi)= \frac{2\pi}{i\omega}\sum_{m=-\infty}^{+\infty}
\sum_{l=|m|}^{+\infty}(-i)^{m}e^{im\varphi} \left \{
S_{lm}^{a\omega}(\theta)S_{lm}^{a\omega}(\gamma)(e^{2i\delta_{lm}} -1)
- Y_{lm}(\theta) Y_{lm}(\gamma) (e^{2i\delta_l^N} -1) \right \}
\label{fDoff}
\end{equation}
The corresponding on-axis expression is,    
\begin{equation}
f_D(\theta)= \frac{1}{2i\omega}\sum_{l=0}^{+\infty}
 \left \{ 4 \pi S_{l0}^{a\omega}(\theta) S_{l0}^{a\omega}(0)
( e^{2i\delta_l} -1) -(2l+1)P_l(\cos\theta) (e^{2i\delta_l^N} -1) \right \}
\label{fDon}
\end{equation}
One would expect the sums in (\ref{fDoff}) and (\ref{fDon}) to converge.
This follows from the fact that for $l/\omega M \to \infty $ we 
have $\delta_{lm}\to \delta_{l}^{N}$ and $ S_{lm}^{a\omega}(\theta) 
e^{im\varphi}
\to Y_{lm}(\theta,\varphi)$ \cite{Handler}. We introduce 
a negligible error by truncating the sums at a large value $l_{\rm max}$ 
(say). In practice, $l_{\rm max}$ need not be very large.  We find that
 a value $\sim 30-50$ for $\omega M \lesssim 2 $ typically
suffices. Since each partial wave can be
labelled by an impact parameter $b(l)$ (see Section~IID), the criterion for
$l_{\max}$ to be a ``good'' choice, is that $ b(l_{\rm max}) \gg  b_{\rm c}$, 
where $ b_{\rm c}$ is the (largest) critical impact parameter associated with
an unstable photon orbit in the Kerr geometry.   
 
The truncation of the partial-wave sums
will introduce interference oscillations in the final cross sections 
(roughly with a wavelength $2\pi/l_{\rm max}$ \cite{Handler}). 
These unphysical oscillations can be eliminated by
following the approach of Handler and Matzner \cite{Handler}. 
For a chosen $\l_{\rm max}$ 
we add a constant $\beta$ to all the phase-shifts in (\ref{fDoff}) and
(\ref{fDon}). This constant is chosen such that
$\delta_{l_{\rm max},m} + \beta = \delta_{l}^{N} $. This means that
the resulting cross section is effectively smoothed.  


\subsection{Deflection functions}

It is well-know that the so-called deflection function is of
prime importance in scattering problems. 
It arises in the semiclassical description
of scattering, as discussed in the pioneering work of Ford and 
Wheeler \cite{Ford}. Although these authors considered 
scattering in the context of quantum theory, their formalism is readily 
extended to the black-hole case.
In the semiclassical paradigm, the phase-shifts are approximated by a 
one-turning point WKB formula (typically useful for $l$ 
much larger than unity).  

In a problem 
which has only one classical turning point, the deflection function 
is defined as
\begin{equation}
\Theta(l)= 2\frac{d\delta^{\rm WKB}}{dl}
\end{equation}
where $l$ is assumed to take on continous real values. As a convention, the 
deflection function is negative for attractive potentials.
The right-hand side of this equation resembles the expression 
for the deflection angle of classical motion in the 
given potential, provided that we define
the following effective impact parameter $b$ for the wave motion \cite{Newton}
\begin{equation}
b= \frac{l+ 1/2}{\omega}
\label{impact}
\end{equation}

The black-hole effective potential has two turning points, but for 
$l/\omega M \to \infty $ the scattering is mainly due to the 
outer turning point and one can derive a one turning point 
WKB approximation for the phase shifts.  
For a Schwarzschild black hole this expression
is \cite{Nils}
\begin{equation}
\delta_{l}^{\rm WKB}= \int_{t}^{+\infty} \left [ Q_{\rm s} - \left (1 -\frac{2M}{r} \right )^{-1} \omega
\right ] dr - \omega t_{\ast} + (2l +1)\frac{\pi}{4}
\label{d_wkb1}
\end{equation} 
where 
\begin{equation}
Q_{\rm s}^2= \left ( 1 -\frac{2M}{r} \right )^{-2} \left [ \omega^2 -
\left ( 1-\frac{2M}{r} \right ) \frac{l(l+1)}{r^2} + \frac{M^2}{r^4} \right ]
\end{equation}
Here  $t_{\ast}$ denote the value of the tortoise coordinate corresponding 
to the (outer) turning point $t$. 
We now define the deflection function as
\begin{equation}
\Theta(l)= 2 Re \left [ \frac{d \delta_{l}}{dl} \right ]
\label{def_sch}
\end{equation}
where only the real part of the phase shift is considered, as the whole
discussion is relevant for elastic scattering only. 
Using (\ref{d_wkb1}), we find
that the real scattering angle is
\begin{equation}
\Theta(l)= \pi -2\left ( \frac{l+1/2}{\omega} \right ) 
\int_{t}^{+\infty} \frac{dr}{r^2} \left [ 1- \left ( 1-\frac{2M}{r} 
\right ) \left (\frac{l+1/2}{\omega r} \right )^2 + {\cal O }(M^2/r^2)
\right ]^{-1/2}
\label{def_sch2} 
\end{equation}
This WKB 
result should be compared to 
the deflection angle for a null geodesic in the Schwarzschild
geometry, which is given by
\begin{equation}
\Theta_c(b)= \pi - 2 b\int_{t}^{+\infty} \frac{dr}{r^2} \left [ 1-
\left ( 1-\frac{2M}{r} \right )\frac{b^2}{r^2} \right ]^{-1/2}
\label{def_cl}
\end{equation}
where $ b= L_{z}/E $ is the orbit's impact parameter ($L_{z}$ and $E$ denote,
respectively, the orbital angular momentum component along the black hole's
spin axis and the orbital energy) and $t$ is the (classical) turning point.
In writing down these expressions we have chosen the signs in such a way 
that the deflection angle is negative for attractive potentials.
Clearly,  it is  possible to ``match'' the deflection function 
(\ref{def_sch}) with the classical deflection angle, albeit only at 
large distances. Since the effective impact parameter will be given
by (\ref{impact}), it is clear that in (\ref{def_sch2}) the integral will be
over large $r$ only.   

Owing to its clear geometrical meaning the deflection function is an 
exceptionally useful tool in scattering theory. It can be used 
to define diffraction phenomena like glories, rainbows etc. \cite{Ford}. For
example, in axisymmetric scattering backward glories are present 
if the deflection function takes on any of the values $\Theta=-n\pi$, where 
$n$ a positive odd integer.       

It seems natural to try and define deflection functions
for Kerr scattering as well. In general, we anticipate the need for 
two deflection functions $\Theta(l,m)$ and $\Phi(l,m)$ (with only the 
first being relevant for the special case of on-axis scattering). 
The WKB  phase-shift formula becomes in the Kerr case: 
\begin{equation}
\delta_{lm}^{\rm WKB}= \int_{t}^{+\infty} \left [ Q_{\rm k} - 
\left ( \frac{r^2 + a^2}{\Delta} \right ) \omega \right ] dr 
- \omega t_{\ast} + (2l +1)\frac{\pi}{4}
\label{d_wkb2}
\end{equation}
where  
\begin{equation}
Q_{\rm k}^2= \frac{1}{\Delta^2} \left [ K^2 - \lambda\Delta + M^2 -a^2 \right ]
\end{equation}
where $\lambda = E_{lm} + a^2 \omega^2 - 2am\omega$.

The next step is to derive the deflection angles for null geodesics approaching
a Kerr black hole from infinity. Such orbits are studied in detail in 
\cite{Chandra}. From the results in \cite{Chandra} it is clear that 
it is not easy to write down a general expression for the deflection angle 
in the Kerr case. But we can obtain useful results in two particular
cases.

We begin by considering a null ray with $L_z=0$ (which would correspond to
an axially incident partial wave). For such an orbit we find that 
the deflection angle $\Theta_{\rm c}$ obeys the 
 following relation
\begin{equation}
\int_{\pi}^{\Theta_{\rm c}(\eta)} d\theta \left [ 1 + \frac{a^2}{\eta^2} 
\cos^2\theta \right ]^{-1/2} = -2\eta\int_{t}^{+\infty} \frac{dr}{r^2}
\left [ 1 -\frac{\eta^2}{r^2} \left (1 -\frac{2M}{r} + \frac{a^2}{r^2} \right )
+ \frac{a^2}{r^2} + \frac{2a^2 M}{r^3} \right ]^{-1/2} 
\label{latdefc}
\end{equation}
where $\eta= C^{1/2}/E$, with $C$ denoting the orbit's Carter constant.
This expression is valid provided the ray's $\theta$-coordinate varies
monotonically during scattering. This should be true in the cases we are interested in, at least for large impact
parameters such that $ \eta \gg M$. 
The ray will also be deflected in the $\varphi$-direction but this deflection
carries no information regarding plane-wave scattering due to the axisymmetry
of the problem.

We next consider a null ray travelling in the black hole's equatorial plane.
This situation will be particularly relevant for a plane wave incident 
along $\gamma=\pi/2$.
The net azimuthal deflection $\Phi_{\rm c}(b)$ for an impact parameter
$b= L_{\rm z}/E$ is
\begin{equation}
\Phi_{\rm c}(b)= \pi - 2 b \int_{t}^{+\infty} \frac{dr}{r}
\left [ 1 - \frac{a^2}{\Delta} + \frac{2aMr}{\Delta b} \right ]
\left [ r^2 + a^2 + \frac{2a^2 M}{r} - b^2 \left ( 1 -\frac{2M}{r} \right)
-\frac{4aMb}{r} \right ]^{-1/2}
\label{azimdefc}
\end{equation}
Working to the same accuracy in terms of $M/r$ as in the Schwarzschild case,
we can match (\ref{latdefc}) and (\ref{azimdefc}) to 
$\partial\delta_{lm}/\partial l$. This matching becomes possible
if we use the following approximate expression for the eigenvalue 
$E_{lm}$ \cite{Handler}
\begin{equation}
E_{lm}\approx l(l+1) - \frac{1}{2}a^2\omega^2 + {\cal O}(\frac{a^3\omega^3}{l})
\end{equation}
As in the Schwarzschild case we assume that
the effective impact parameter is given
by (\ref{impact}). Although there is no
occurrence of the multipole $m$ in the above expressions, one can 
argue (from the symmetry of the various spheroidal harmonics, which is
similar to that of the spherical harmonic of the same ($l,m)$) 
that the classical angles
(\ref{latdefc}) and (\ref{azimdefc}) are related to partial
waves with $m=0$ and incidence $\gamma=0$ and  partial waves with 
$m= \pm l$ and incidence $\gamma=\pi/2$, respectively.
Hence, we define the latitudinal deflection function   
\begin{equation}
\Theta(l) = 2Re \left [ \frac{\partial\delta_{lm}}{\partial l}(m=0) \right ]
\label{latdef}
\end{equation}
and the azimuthal (``equatorial'') deflection function 
\begin{equation}
\Phi(l)= 2Re \left [ \frac{\partial\delta_{lm}}{\partial l}(m=\pm l) \right ]
\label{azimdef}
\end{equation}


\section{Numerical results}

\subsection{Phase-shifts calculation via the Pr\"{u}fer transformation}

In order to determine the required scattering 
phase-shifts we have used a
slightly modified version of the simple Pr\"{u}fer transformation, 
well-known from the numerical analysis of Sturm-Liouville problems 
\cite{Pryce}. The method is best illustrated by a standard
second order ordinary differential equation:
\begin{equation}
\frac{d^2 \psi}{dx^2} + U(x) \psi=0
\label{sle}
\end{equation}
where we can think of $x$ as being a radial coordinate, spanning the entire
real axis, and $U$  an effective potential (in our problem 
corresponding to a single potential barrier) with asymptotic behaviour
\begin{equation}
U(x) \sim \left\{ \begin{array} {ll}
k^2 \quad \mbox{as } x  \to -\infty \ , \\ 
\omega^2 \quad \mbox{as } x \to +\infty \ .
\end{array} \right.
\end{equation} 
with $\omega$ and $k$  real constants. (The black hole problem we are
interested in does, of course, have exactly this nature.)
The solution of
(\ref{sle}) will take the form of oscillating 
exponentials for $x \to \pm \infty$.
Let us assume that we are looking for a solution to (\ref{sle}) with
purely ``ingoing'' behaviour at the ``left'' boundary ($x \to - \infty$)
and mixed ingoing/outgoing behaviour as $x \to + \infty$: 
\begin{equation}
\psi \sim \left\{ \begin{array}{ll}
e^{-ikx} \quad \mbox{as } x  \to -\infty \ , \\ 
B \sin[\omega x + \zeta ] \quad\mbox{as } x \to +\infty \ .
\end{array} \right.
\end{equation}
where $\zeta$ and $B$ are complex constants. We can then write the exact solution of (\ref{sle}) in the form 
\begin{equation}
\psi(x)= e^{\int P(x) dx} 
\end{equation}
The function $P(x)$ is the logarithmic derivative of $\psi(x)$ 
(a prime denotes derivative with respect to $x$)
\begin{equation}
\frac{\psi^{\prime}}{\psi}= P
\end{equation}
which obeys the boundary condition $P(x) \to -ik$ for $x \to -\infty$. 

Similarly, we can express the function $\psi$ and its 
derivative via a Pr\"{u}fer transformation; 
\begin{eqnarray}
\psi(x) &=& B\sin[\omega x +\tilde{P}(x)] 
\\
\psi^{\prime}(x)&=&B \omega \cos[\omega x + \tilde{P}(x) ]
\end{eqnarray}
with $\tilde{P}(x)$ a Pr\"{u}fer phase function which has $\zeta$ as 
its limiting value for $x \to + \infty$. Direct substitution in (\ref{sle}) yields the 
equations
\begin{equation}
\frac{dP}{dx} + P^2 + U(x) = 0
\label{P}
\end{equation}
\begin{equation}
\frac{d\tilde{P}}{dx} + \left [ \omega -\frac{U(x)}{\omega} \right ]
\sin^2(\tilde{P} + \omega x)= 0
\label{P2}
\end{equation} 
The idea is to numerically integrate (\ref{P}) and (\ref{P2}) instead
of the original equation (\ref{sle}). 
The motivation for this is that, while the original solution may 
be rapidly oscillating, the phase-functions $P$ and $\tilde{P}$
are expected to be slowly varying functions of $x$.

We expect this integration scheme to be considerably more stable, 
especially for high frequencies, than any direct approach to (\ref{sle}). 
Moreover, eqs. (\ref{P}) and (\ref{P2}) are well behaved also at the 
classical turning points and are well suited for barrier 
penetration problems.  However, if we want to ensure that the 
phase-functions are smooth and non-oscillatory we must account for the 
so-called Stokes phenomenon --- the switching on of small exponentials in 
the solution to an equation 
of form (\ref{sle}). To do this we simply shift from studying $P(x)$  
(which is calculated from $ x=-\infty $ up to the relevant 
matching point $x_{\rm m}$) to  $\tilde{P}(x)$  
(which is calculated outwards to $x=+\infty$). 
In practice, the calculation is stable and reliable if the switch is
done in the vicinity of the maximum of the black-hole potential 
barrier (the essential key is to not use one single 
representation of the solution through the entire potential barrier). 
The two phase functions are easily connected by      
\begin{equation}
\tilde{P}(x)= -\omega x + \frac{1}{2i} \ln\left[ \frac{iP -\omega}
{iP +\omega} \right ]
\end{equation}
Finally, the desired phase-shift can be easily extracted as 
$\delta_{lm}= \zeta + l\pi/2$.

A major advantage of the adopted method is that it permits 
direct calculation of the partial derivatives 
 $\partial \delta_{lm}/\partial l~$ and  
$\partial\delta_{lm}/\partial m$, which are required for the
evaluation of the  
deflection functions from Section~IID. 
The equations for the $l,m$-derivatives of 
$P$ and $\tilde{P}$ are simply found by 
differentiation of (\ref{P}) and  (\ref{P2}).  
For the case of scattering by a Kerr black hole, calculation of 
$\delta_{ lm}$ and its derivatives with respect to $l,m$ requires 
knowledge of the angular eigenvalue $E_{lm}$ (see Appendix~C)
and its $l,m$-derivatives. 
We have used an approximate formula which is a polynomial expansion in 
$a\omega$ (formula $21.7.5$ of \cite{Stegun}). 
This expression (and its derivatives) is well behaved 
for all integer values of $l$ and $m$. 
However, it is divergent for the half-integer values $ l= 1/2, 3/2, 5/2 $.
Hence, the numerical calculation of the  deflection function will fail 
at these points, and will be generally ill-behaved in their neighbourhood. 
For the full cross section calculation, we additionally need to calculate
the spin-0 spheroidal harmonics $S_{lm}^{a\omega}(\theta)$. This 
calculation is discussed in detail in Appendix~C.


\subsection{Schwarzschild results}

In this section we reproduce phase-shifts and cross sections for 
 Schwarzschild scattering. The purpose of this exercise is to 
validate, and demonstrate the reliability
of, our numerical methods. We  compare our numerical
integration results  
to ones obtained using the phase-integral method \cite{Nils}.

\begin{figure}[tbh]
\centerline{ \epsfysize=4cm \epsfbox{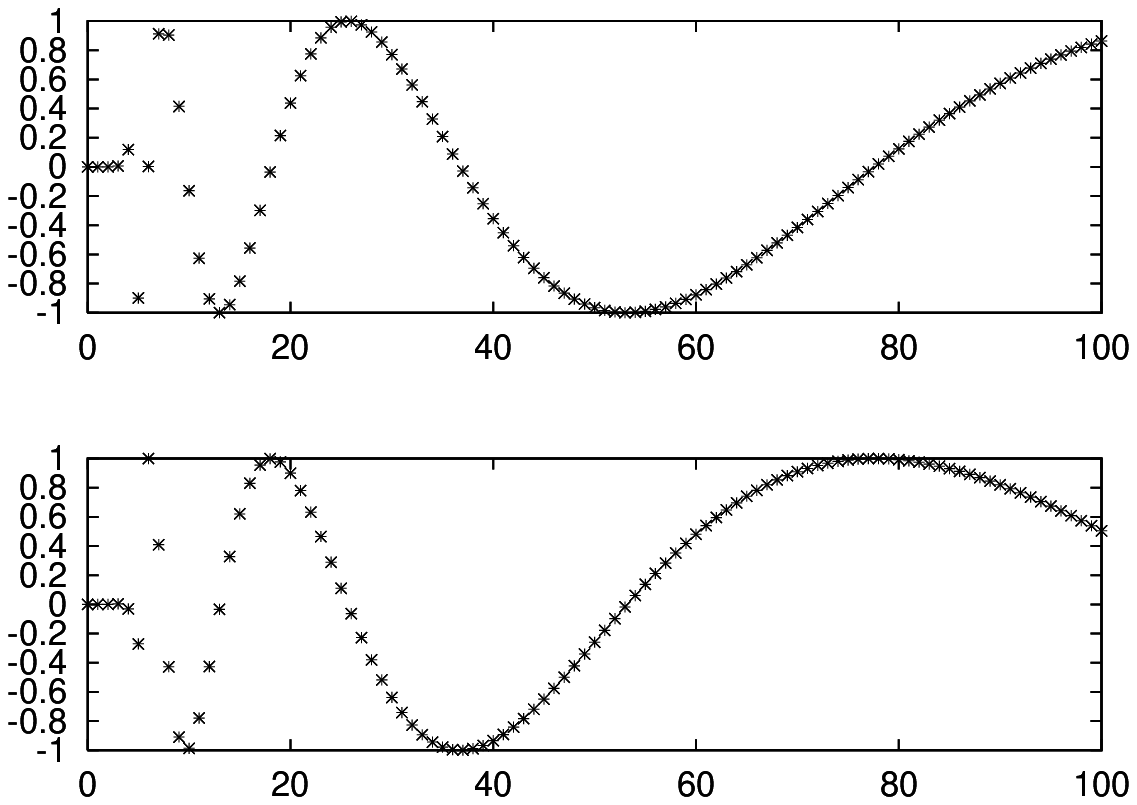}}
\vspace{4cm} 
\caption{Comparison of numerical phase-shifts for a Schwarzschild 
black hole 
(cross), against phase-integral data (plus). 
We show the real (upper frame) and imaginary (lower frame) 
parts of the scattering matrix element $ {\cal S}_{l}= e^{2i\delta_l}$ 
 as functions of $l$.
The agreement between the two methods is clearly excellent.}
\label{fig1}
\end{figure}

As a first crucial test we compare, in Fig.~\ref{fig1}, 
the first 100 phase-shifts for $\omega M=1$. 
Because of the 
multi-valued nature of the phase-shifts we always plot the 
quantity ${\cal S}_l=e^{2i\delta_l}$.
As is evident from Fig.~\ref{fig1}, 
the agreement between our numerical phase-shifts
and the phase-integral ones is excellent. This is equally true for a 
all frequencies examined (up to $\omega M=10$). 
It should be noted that ${\cal S}_l$ is essentially zero 
($\delta_l$ has a positive imaginary part) for those values of $l$ 
for which absorption by the black hole is important. 
That this is the case for the lowest multipoles is clear from Fig.~\ref{fig1}.
As $l$ increases $\delta_l$ becomes almost real and
as a consequence ${\cal S}_l$ is almost purely oscillating. 

\begin{figure}[tbh]
\centerline{\epsfysize=7cm \epsfbox{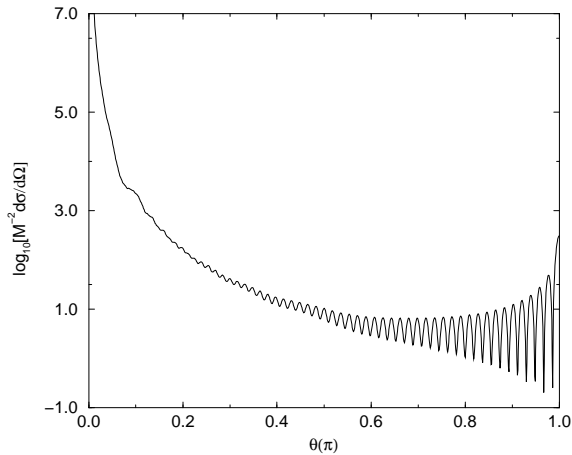}} 
\caption{Differential cross section for scattering of a wave 
with relatively high frequency, $\omega M=10$, 
from a Schwarzschild black hole, based on the 
the first 200 partial wave phase-shifts. The backward glory oscillations are 
prominent. }
\label{fig2}
\end{figure}

In Fig.~\ref{fig2} we present an $\omega M=10$ cross section 
generated from our numerical phase-shifts. For this particular 
calculation we have used $l_{\rm max}= 200$. The resulting cross 
section  matches the one constructed using 
approximate phase-integral phase shifts perfectly. This  
demonstrates the efficiency of our approach in the high frequency regime, and
it is clear that our study of Kerr scattering will not be limited 
by the lack of reliable phase shifts. However, the Kerr study is 
nevertheless limited in the sense that $\omega$ cannot be taken to be 
arbitrarily large. This restriction is  imposed by the
calculation of the spheroidal harmonics (Appendix C).  
However, it is important to emphasize that the most interesting
frequency range, as far as diffraction phenomena is concerned, is
$\omega M \sim 2$ \cite{Nils}. Thus, we expect that our investigation 
should be able to reliable unveil all relevant rotational 
effects in the scattering problem.
  
We also find that 
the numerical values for  $d\delta_l/dl$ are in good
agreement with the respective phase-integral results.
As a final remark we should emphasize that the numerical approach adopted in
this work is very efficient from a computational point of view. 



\subsection{On-axis Kerr scattering}

\begin{figure}[tbh]
\centerline{ \epsfysize=5cm \epsfbox{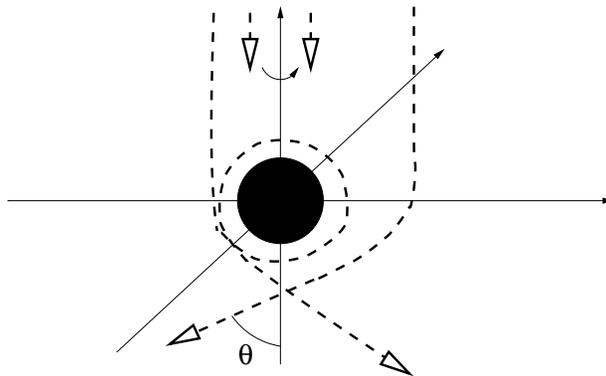}}
\caption{A schematic drawing illustrating the case of on-axis 
scattering from a rotating black hole. Because of the axial symmetry 
of the problem, the scattering results are qualitatively 
similar to those for a Schwarzschild black hole. Two rays, which emerge
having been scattered by the same angle $\theta$ are indicated. }
\label{draw2}\end{figure}

Having confirmed the reliability of our numerical results 
we now turn to the study of scattering
of axially incident scalar waves by a Kerr black hole, cf. Fig.~\ref{draw2}. 
In principle, we would  expect
the corresponding cross sections to be qualitatively
similar to the Schwarzschild ones. 
The main reason for this is the inability of axially
impinging partial waves to distinguish between prograde and retrograde
orbits.  However, examination of the orbital equations \cite{Chandra} reveals
that the critical impact parameter (associated with the unstable photon orbit)
decrease sligthly from the value
$3\sqrt{3}M $ as the black hole spins up. For example, for a=0.99M we have
$b_{c}= 4.74M$. 

\begin{figure}[tbh]
\centerline{\epsfysize=7cm \epsfbox{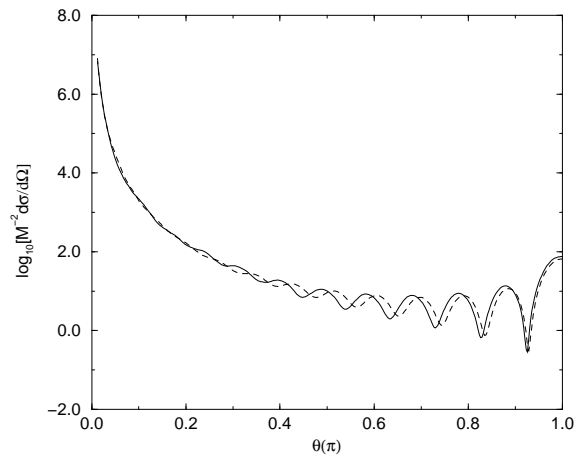}} 
\caption{Differential cross section for on-axis scattering from a 
Kerr black hole (solid line). The black
hole's spin is $a=0.99M$ and $\omega M=2$. 
The graph is based on data for $l_{max}=30$.
 The dashed line represents the corresponding Schwarzschild cross
section.}
\label{figoncs}
\end{figure}

Indeed, Fig.~\ref{figoncs} confirms our expectations. 
The data in the figure corresponds to
a black hole  with spin $a=0.99M$ and a wave frequency of $\omega M=2$.
For purposes of comparison, we also show the corresponding 
Schwarzschild cross section. 
The two cross sections are very similar. In particular, 
they are both dominated by the backward glory. 
It is well-known \cite{book,path_int1} that for 
Schwarzschild scattering the 
glory effect can be described in terms of Bessel functions. 
It has been shown that for $\theta \approx \pi$ the glory cross 
section can be approximated by
\begin{equation}
\frac{d\sigma}{d\Omega}(\theta)|_{\rm glory} \propto J^{2}_{0}[\omega
 b_{\rm c} \sin\theta]
\label{glory}
\end{equation} 
A similar result holds for the backward glory in the case of on-axis
Kerr scattering. Since $b_c$ gets smaller, one would expect the zeros 
of the Bessel
function (the diffraction minima) 
to move further away from $\theta = \pi$ as $a$ increases. 
This effect is indicated by the data in Fig.~\ref{figoncs}.   

\begin{figure}[tbh]
\centerline{\epsfxsize=10cm \epsfysize=7cm \epsfbox{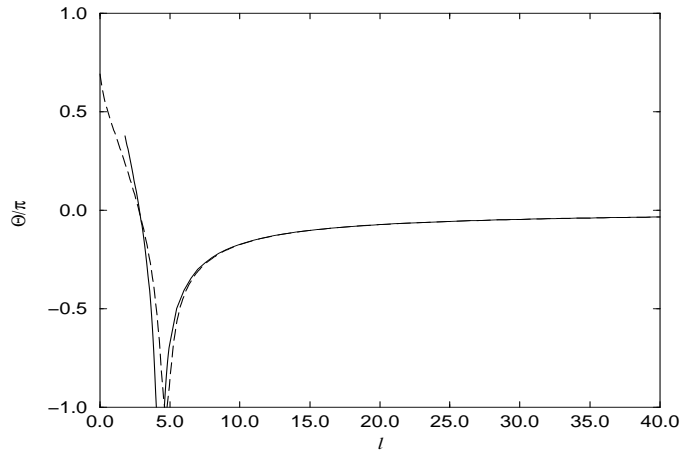}} 
\caption{Deflection function (in units of $\pi$) for on-axis scattering
from a Kerr black hole. 
The black hole spin is $a= 0.99M$ and $\omega M=1 $. 
The dashed curve is the corresponding Schwarzschild deflection function. 
This figure confirms the
behaviour expected from the geometric optics considerations, namely, 
the slight decrease of the critical impact parameter with increasing $a$.}
\label{figThet}
\end{figure}

The above conclusions are further supported 
by the results for the  deflection function 
(\ref{latdef}), as shown in Fig.~\ref{figThet}. 
Because of the inaccuracies inherent in our method of 
calculating the deflection function for the lowest $l$-multipoles, 
see the discussion in
Section IIIA, we do not show results for this regime.
This is, however, irrelevant as the corresponding partial 
waves are expected to be more or less completely
absorbed by the black hole.  

\subsection{Off-axis Kerr scattering}

\begin{figure}[tbh]
\centerline{ \epsfysize=5cm \epsfbox{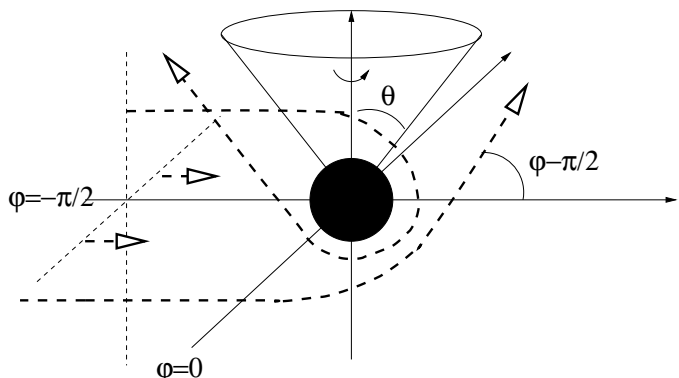}}
\caption{A schematic drawing illustrating  off-axis 
scattering from a rotating black hole. We show (as thick dashed lines)
two ``rays'', one of which corresponds to motion in the
black hole's equatorial plane. }
\label{draw3}\end{figure}

The  conclusions of our study of on-axis scattering are perhaps  
not very exciting. Once the Schwarzschild case is understood, the 
on-axis results for Kerr come as no surprise. This is, however, not
the case for off-axis scattering, cf. Fig.~\ref{draw3}, 
where several new  features appear.

Our study of the off-axis case
provides the first results for non-axisymmetric wave
scattering in black hole physics. 
Since this problem has not been discussed in great detail previously, it is 
worthwhile asking whether we can 
make any predictions before turning to the numerical calculations. 
Two effects ought to be relevant:
First of all, the partial waves  now
have orbital angular momentum  which couples to the black hole's spin. 
As a result the partial waves can be divided into 
prograde ($m >0$) and retrograde ($m<0$) ones.
We expect prograde waves to be able to approach closer to the horizon
than retrograde ones. In the geometric optics limit, prograde and
retrograde rays tend to have increasingly different critical impact parameters
as $a \to M$.  As a second feature, we expect to find that large $l$ partial
waves will effectively feel only the spherically symmetric 
(Newtonian) gravitational
potential. In other words, partial waves with the same (large) $l$ and
different values of $m$ will approximately acquire the same phase-shift.     

\begin{figure}[tbh]
\centerline{ \epsfysize=7cm \epsfbox{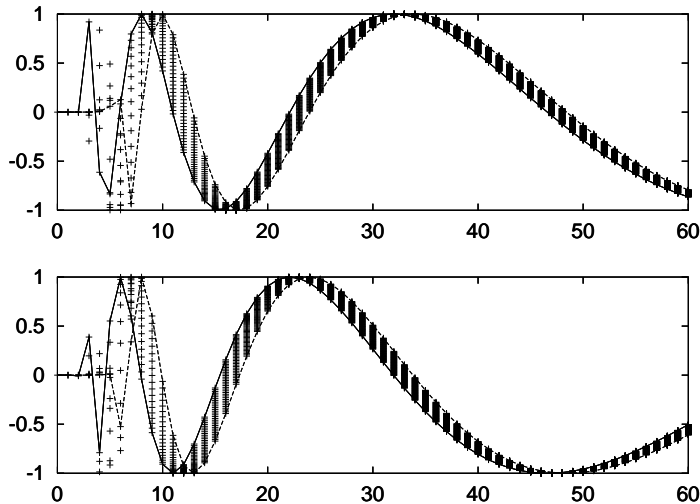}} 
\caption{Off-axis phase-shifts for $\omega M=1$ and $a=0.9M$. 
We illustrate the real (upper panel) and imaginary (lower panel) parts 
of ${\cal S}_{lm}= e^{2i\delta_{lm}}$  as  functions of $l$, including all
permissible values of $m$. The two curves correspond to $m=l$ (solid line) 
and $m=-l$ (dashed line).}
\label{figofps}   
\end{figure}

Our numerical results essentially confirm these expectations, 
as is clear from the phase-shifts (calculated for $a=0.9M$ and 
$\omega M=1$) shown in  Fig.~\ref{figofps}. As above, 
we have graphed the single-valued quantity ${\cal S}_{lm}=e^{2i\delta_{lm}}$
as a function of $l$. For each value of $l$ we have included all the
phase-shifts for $ -l \leq m \leq +l$. The solid (dashed) line corresponds
to $m=+l$ ($m=-l$) and the intermediate values of $m$ lead to results in 
between these two extremes. For $l \gg 1$, 
partial waves with different values of $m$
have almost the same phase-shift. This is easy to deduce from the fact that 
the two curves approach each other as $l$ increases. On the other hand, for 
the first ten or so partial waves we get very different results for
the various values of $m$. In particular,  we see that phase-shifts with $m>0$
become almost real (that is, $| {\cal S}_{lm}|$ becomes non-zero)
for a smaller $l$-value as compared to the $m<0$ ones.
As anticipated, this is due to the different critical impact parameters
associated with prograde/retrograde motion, and the fact that 
a larger number of prograde partial waves are absorbed by the black hole. 


We now turn to the cross section results for the off-axis case. 
We have considered a plane wave
incident along the direction $\gamma=\varphi=\pi/2$. 
Even though our formalism  allows incidence from any direction
we have focussed on this case, which is illustrated in Fig.~\ref{draw3}.
The motivation for this is that there will then be partial waves 
(specifically the ones with $m=\pm l$) that are 
mainly travelling in the black hole's equatorial plane. These partial
waves are important because one would expect them to 
experience the strongest rotational effects. 
Besides, we can obtain an understanding of these
waves by studying equatorial null geodesics in the geometric optics limit.
Equatorial null rays are much easier to describe than 
nonequatorial ones. This proves valuable in attempts to 
``decipher'' the off-axis cross sections, and the obtained conclusions 
provide an understanding also of the general case.

\begin{figure}[tbh]
\centerline{ \epsfysize=8cm \epsfbox{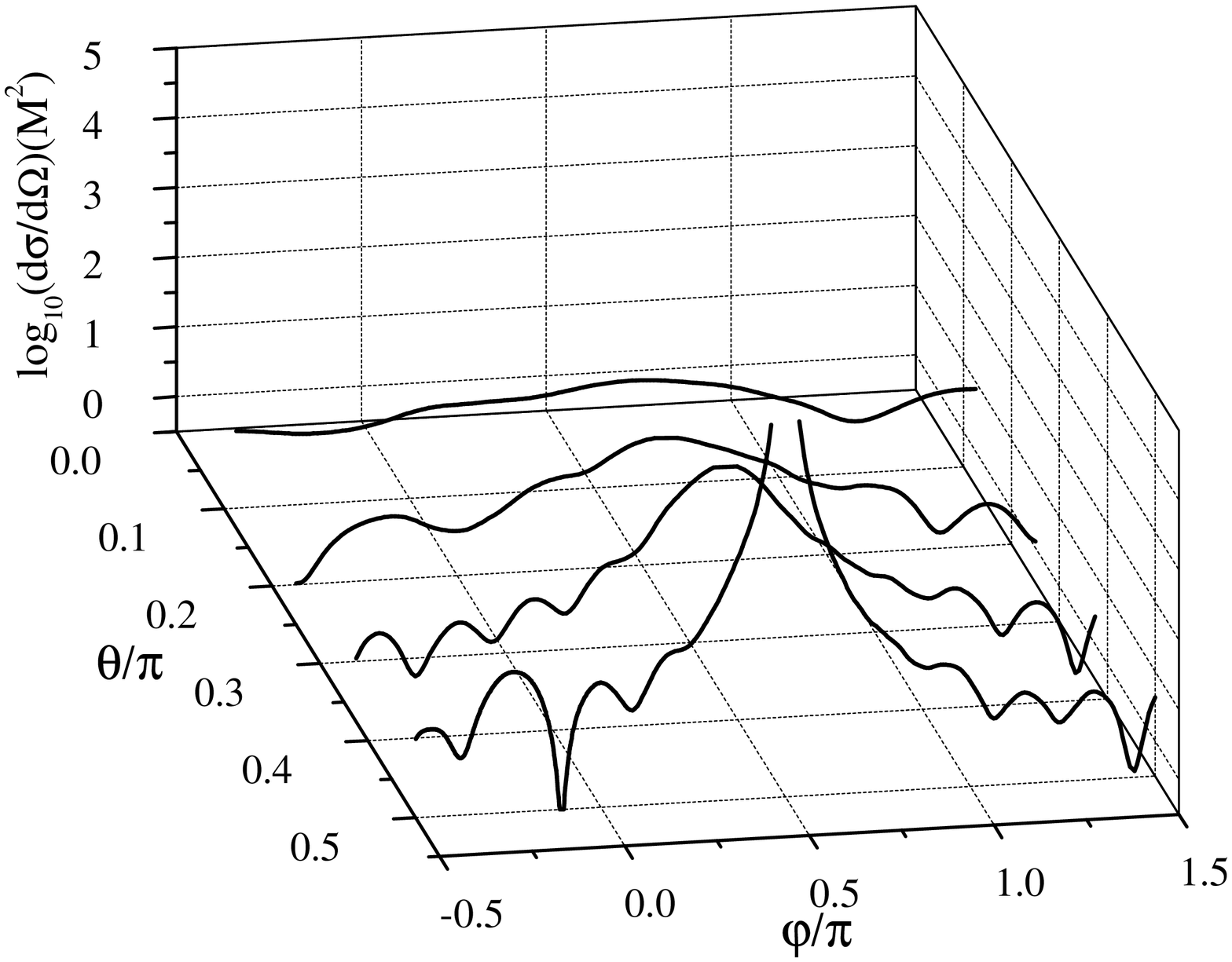}}
\centerline{ \epsfysize=8cm 
\epsfbox{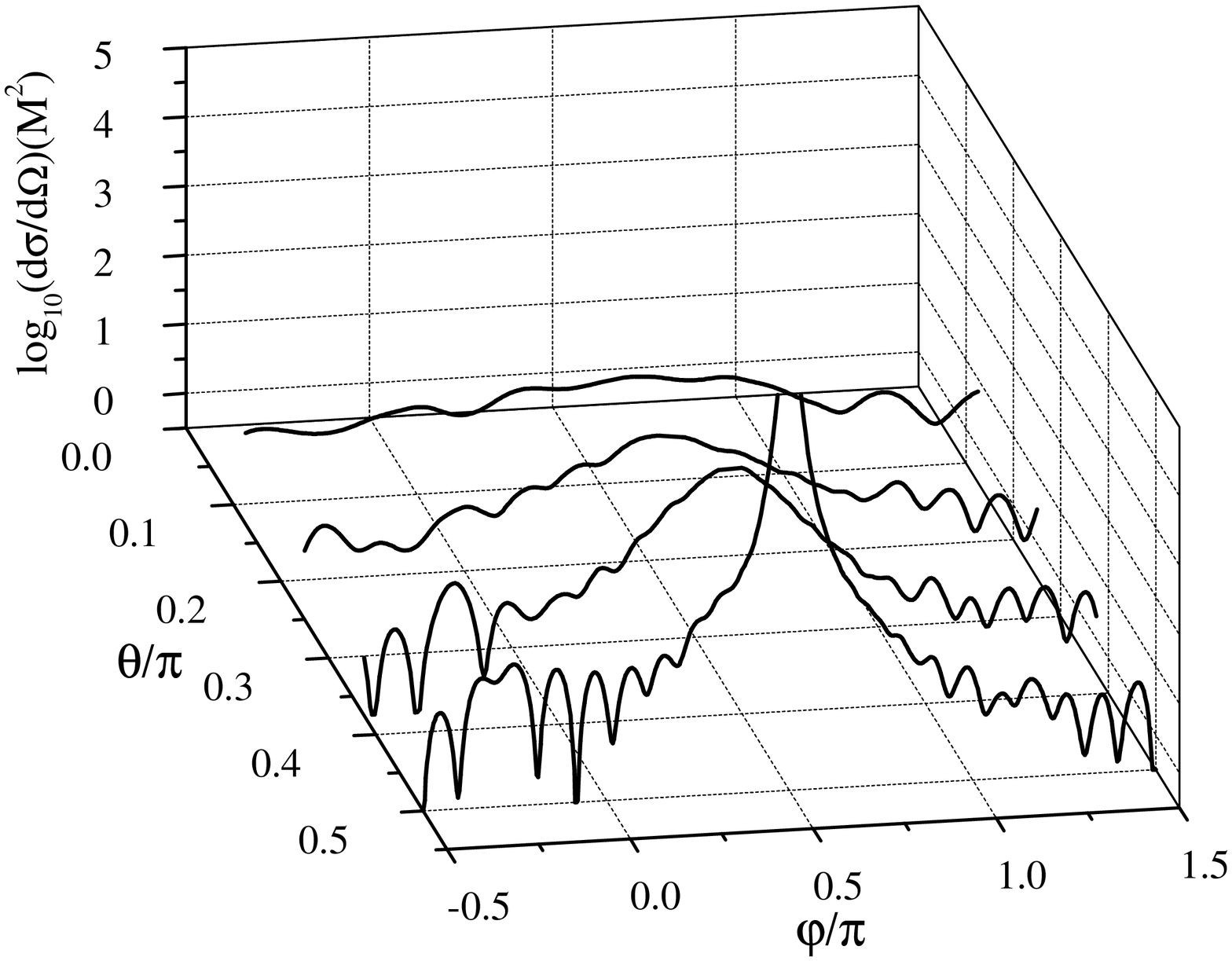}}
\caption{Off-axis cross sections ($\theta= \pi/8, \pi/4, 3\pi/8, \pi/2$) 
for a black hole with spin $a=0.9M$ and scattered waves with frequency
$\omega M=1$ (upper panel) and $\omega M=2$ 
(lower panel). The incident wave is travelling  in the $(\theta,\varphi)=
(\pi/2, \pi/2)$ direction.}
\label{fig6}

\end{figure}

In Fig.~\ref{fig6} we present a series of cross sections as functions of 
$\varphi$  for the specific values $\theta= \pi/8, \pi/4, 3\pi/8, \pi/2 $.
These results correspond to viewing the scattered wave on the circumference
of cones (like that shown in Fig.~\ref{draw3}) with 
increasing opening angles.
Two different frequencies $\omega M=1 $ and $\omega M=2$  
have been considered for a black hole with spin $a=0.9M$.
A first general remark concerns the asymmetry of the cross sections with 
respect to the incidence direction (note, however, that as a consequence 
of our particular choice of
incidence direction there is still a
reflection symmetry with respect to the equator). We can also easily 
distinguish 
the Coulomb forward divergence in the direction $\theta=\varphi=\pi/2$.
Another obvious feature in Fig.~\ref{fig6} 
is the markedly different appearance of the cross
section for different values of
$\theta$. As we move away from the
equatorial plane the cross sections becomes increasingly featureless.
This behaviour is artificial in the sense that as $\theta$ decreases, 
we effectively observe along a smaller circumference. At $\theta=0$ this
circumference degenerates into a point, cf. Fig.~\ref{draw3}.

\begin{figure}[tbh]
\centerline{\epsfysize=8cm  \epsfbox{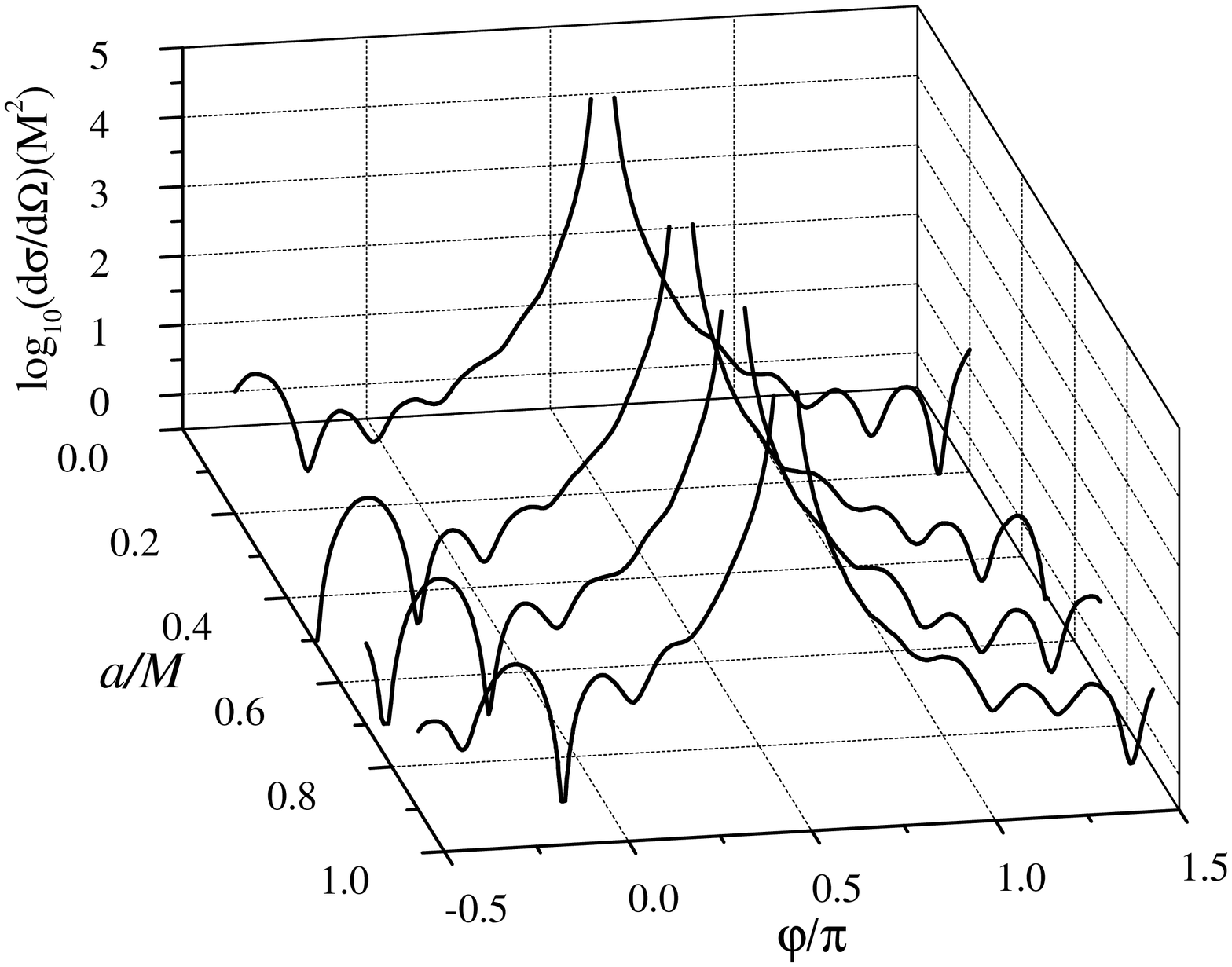}}
\centerline{ \epsfysize=8cm \epsfbox{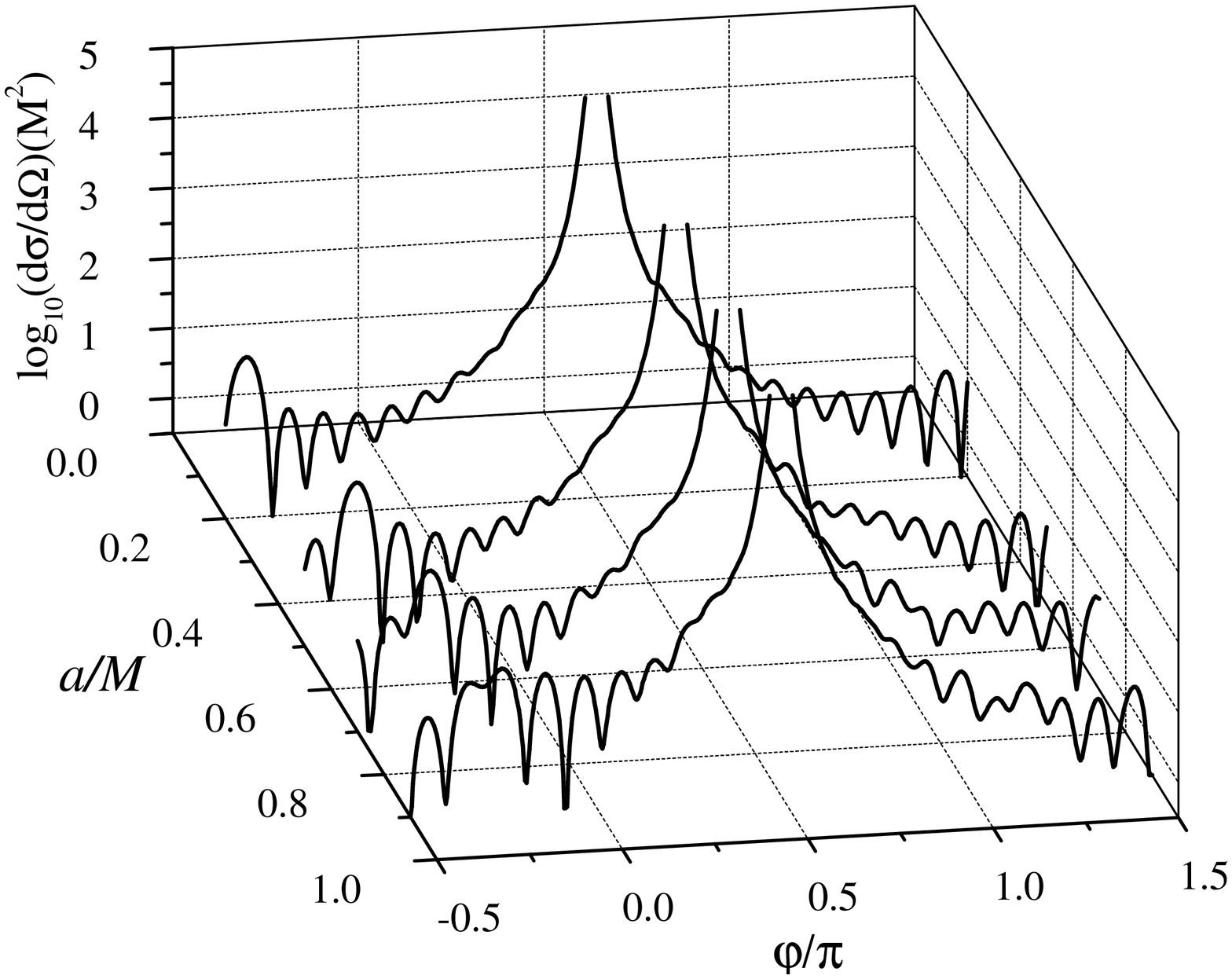}} 
\caption{Off-axis cross sections for $\theta=\pi/2$ and various black-hole 
spins 
(a/M= 0.2, 0.5, 0.7, 0.9) and scattered wave frequencies
$\omega M=1$ (upper panel) and
 $\omega M=2$ (lower panel).}
\label{var_a}
\end{figure}

\begin{figure}[tbh]
\centerline{ \epsfysize=5cm \epsfbox{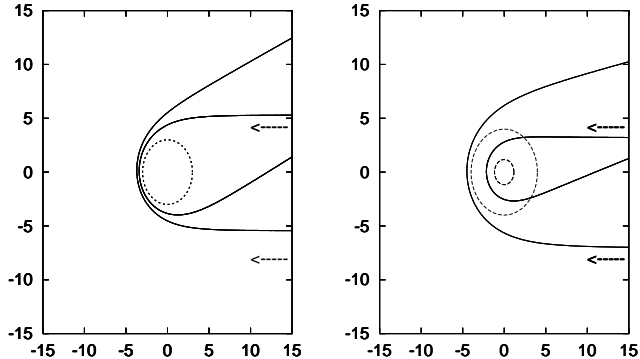}}
\caption{Equatorial null geodesics (viewed from ``above'') 
around a Schwarzschild (left panel) and Kerr black
hole (right panel). The figures are scaled in units of $M$. The rays are 
assumed to arrive from infinity 
(they enter from the right side of each figure in the direction 
indicated by the arrows) in 
parallel directions and exit at the same angle after being scattered. 
The dashed circles represent the unstable photon circular orbits. 
The Kerr black
hole, in the right panel, is taken to rotate counter-clockwise with $a=0.99M$.}
\label{fignulls}
\end{figure}

In order to understand the features seen in Fig.~\ref{fig6} further, we 
 focus  on the $\theta=\pi/2$ cross section. In Fig.~\ref{var_a} 
we show these ``equatorial'' cross sections for a sequence of spin
rates $a/M=  0.2, 0.5, 0.7, 0.9 $. As before, we have considered 
two different wave frequencies, $\omega M=1$ and 
$\omega M=2$. From the results shown in Fig.~\ref{var_a}
it is clear that that the glory maximum is typically not
observed in the backward ($\varphi= -\pi/2$) direction.  
In fact, it is clear that the maximum of the glory oscillations 
move away from the backward direction as the spin of the black hole is 
increased. A similar shift is seen in all  interference oscillations.
This behaviour is easy to explain in terms of the
anticipated rotational frame-dragging.  In order to illustrate this argument,
we consider the geometric optics limit where partial waves are represented
by null rays. Recall that in axisymmetric scattering the backward glory is
associated with the divergence of the classical cross section at $\Theta=\pi$
in such a way that
\begin{equation}
\left (\frac{d\sigma}{d\Omega}\right )_{\rm cl}= \frac{b}{\sin\Theta}
\left (\frac{d\Theta}{db} \right )^{-1}
\end{equation}
The  divergence is a result of the intersection of an infinite number of rays.
For simplicity, let us consider rays travelling in a specified plane. 
Scattering near the backward direction by a Schwarzschild black hole is
illustrated in Fig.~\ref{fignulls} (left panel). Two rays with 
different impact parameters emerge at any given angle. These two waves
make the main contribution to  cross section at that particular angle. 
It is clear that for 
scattering at $\theta=\pi$ the two rays in Fig.~\ref{fignulls} will
follow symmetric trajectories. This means that when observed at infinity, 
after being scattered, the two waves will have equal phases (provided their initial phases were equal). In effect, these two rays will then
constructively interfere in the exact backward direction. 
As we move away from the backward direction, we should observe a series of 
interference maxima and minima --- the two rays will now have an 
overall phase difference since they follow different orbits (see
Fig.~\ref{fignulls}).  
A very crude estimate of the location of the successive maxima would be 
$\theta_{\rm n} \sim n\pi/3\omega M$ where $n=0,1,2, ...$,  
in reasonable agreement with the exact results.

Similar arguments apply in the  case of a Kerr black hole. We shall consider 
only equatorial rays, cf. Fig~\ref{fignulls} (right panel).
As a result of the discrimination between
prograde and retrograde orbits, the 
two rays contributing to the cross section in the exact backward 
direction will no longer
follow symmetric paths. In fact,  the ray symmetric to the prograde
ray shown in Fig~\ref{fignulls} will follow a plunging orbit. 
Therefore, we should not expect the interference maximum to be located in 
the exact backward direction. 
An estimate (based on a crude calculation of the phase 
difference between the two null rays) of the location of the main 
backward glory maximum yields
\begin{equation}
 \varphi_{\rm max} \sim \pi \left ( \frac{r_{\rm ph+} -r_{\rm ph-} }
{r_{\rm ph+} +r_{\rm ph-}} \right )
\end{equation}
where $ r_{\rm ph+}$ and $ r_{\rm ph-}$ denote, respectively, the location of
the prograde and 
retrograde unstable photon orbits (in Boyer-Lindquist coordinates). 
This angle is measured from the backward direction in the direction of 
the black hole's rotation. This simple prediction agrees 
reasonably well with the results inferred from
our numerical cross sections. In a similar way, all other 
maxima and minima will be frame-dragged in the black hole's 
rotational direction.       

\begin{figure}[tbh]
\centerline{ \epsfysize=6cm \epsfbox{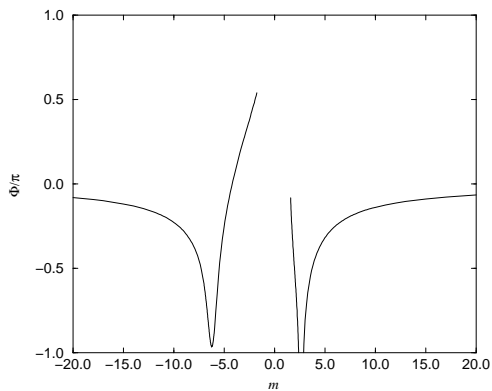}} 
\caption{The deflection function $\Phi$ for off-axis scattering 
is shown as a function of $m$ for ``equatorial'' partial waves $m= \pm l$. 
The black hole spin is $a=0.9M$ and the wave frequency is $\omega M=1 $.
The small $|m|$ region is not included because of the inaccuracies 
discussed in Section~IIIA.
}
\label{figPhi}
\end{figure}

To complete this discussion we 
consider the deflection function $\Phi(l,m)$ for ``equatorial''
partial waves ($m=\pm l$), an $a=0.9M$ black hole and $\omega M=1$. The
corresponding data is shown in Fig.~\ref{figPhi}. 
There are two distinct logarithmic divergences which are 
associated with the existence of separate unstable circular photon orbits 
for prograde and retrograde motion. Note that for $m>0$ the 
deflection function diverges steeper than it does for $m<0$. The origin of this
effect is the fact that prograde partial waves with $ b \sim b_{\rm c}$ 
perform a greater number of revolutions
(before escaping to infinity) than retrograde ones. 
Finally, for $ |m|\gg 1$ we recover, as expected, the 
Einstein deflection angle $\Phi \approx -4M/b $
(not explicitly shown in the figure). 

\subsection{Digression: forward glories}

The results presented in the preceeding sections clearly show that, 
in general, black hole cross sections are dominated by a 
``Coulomb divergence'' in the forward direction and (frame-dragged) glory 
oscillations near the backward direction.
However, according to the predictions of 
geometrical optics \cite{book}, one would
expect to find glory oscillations also in the forward direction (see 
comments in \cite{Nils}). For the case of Schwarzschild scattering,
this effect would be associated with partial waves scattered at angles $\Theta=
0, -2\pi, -4\pi, ...$. Inspection of the relevant deflection function
(Fig.~\ref{figThet}) indicates that a partial wave which has $\Theta=0$ 
will also be
strongly absorbed, since it has an impact parameter $b < b_{\rm c}$.
Hence, we would expect its contribution to the forward glory to be 
severely supressed. It thus follows that, as far as the possible forward 
glory is concerned, the most important partial
waves are those with $\Theta= -2\pi$.
These partial waves  ``whirl'' around the black hole as they have
$b \approx b_{\rm c}$.

Ford and Wheeler's semiclassical approach \cite{Ford} shows that
the forward glory is well approximated (for $\theta \approx 0$) by
(\ref{glory}), although with a slightly different proportionality factor.
However, we should obviously not expect to see a pronounced forward 
glory in the cross section, as it will drown in the forward Coulomb
divergence. Still, as an experiment aimed at supporting our intuition, we
can try to ``dig out'' the forward glory pattern. This has to be done 
in a somewhat
artificial manner, but since  the forward glory
is due to scattering and interference of partial waves with 
$b \approx b_{\rm c} $ we can isolate their contribution by truncating the
sum in $f_D$ at some $l_{\rm max} \sim \omega b_{\rm c} $ and at the 
same time neglecting the Newtonian part $f_N$ entirely. 
It is, of course, important to realize that this
``truncated'' cross section is not a physical (observable) quantity. 

\begin{figure}[tbh]
\centerline{\epsfxsize=8cm \epsfysize=7cm \epsfbox{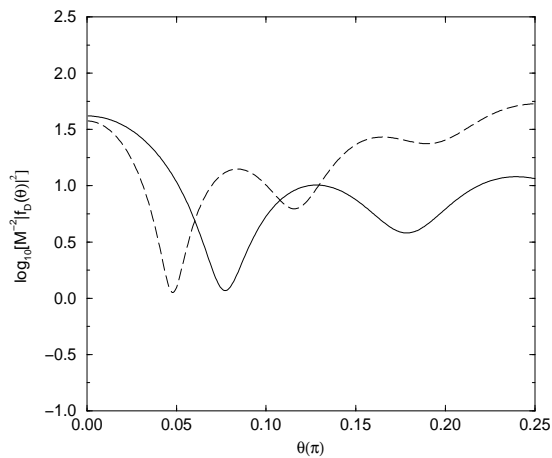}} 
\caption{Illustration of a forward glory in Schwarzschild scattering.
The ``diffraction'' piece $|f_D(\theta)|^2$ of the cross section is 
shown in the vicinity of the forward direction, for wave frequency $\omega M=2$ and for $l_{\rm max}=10$ (solid curve) and $l_{\rm max}=15$ (dashed curve).}
\label{Sfglor}
\end{figure}

In Fig.~\ref{Sfglor} we show the result of this ``truncated
cross section'' calculation for 
the case of a 
Schwarzschild black hole and $\omega M=2 $. We compare results
for two levels of truncation,
$l_{\rm max}= 10 $ and $15 $. In the first case, a clear Bessel-
function like
behaviour arises (it is straightforward to fit a $ J_{0}^2(\omega 
b_{\rm c} \sin\theta) $ function to the solid curve in Fig.~\ref{Sfglor}).
This confirms our expectation that there is, indeed, a forward 
glory present in the data. 
As more partial waves are included the cross section 
begins to  deviate from the glory behaviour, and if we increase
$l_{\rm max}$ further the forward glory is swamped by terms 
that contribute to the Coulomb divergence.

\subsection{The role of superradiance in the scattering of 
monochromatic waves.}

Superradiance is an
interesting effect known to be relevant for
rotating black holes. It is easily understood from the 
asymptotic behaviour (\ref{uin}) of the causal solution to the 
scalar-field Teukolsky equation (\ref{Teuk}). 
If we use this solution and its complex conjugate, and the fact that
two linearly independent solutions to (\ref{Teuk}) must lead
to a constant Wronskian, it is not difficult to show that
\begin{equation} 
( 1 - m \omega_+/\omega ) |{\cal T}_{lm}|^2  = 1 - |{\cal S}_{lm}|^2  \ .
\end{equation}
where we have defined
\begin{equation}
|{\cal T}_{lm}|^2 = \left| {1 \over A^{\rm in}_{lm} } \right|^2  \ .
\end{equation}
From the above result it is evident that the scattered waves are amplified
($|{\cal S}_{lm}|^2>1$) if $\omega < m\,\omega_+$. 
This amplification is known as superradiance.

In principle, one would expect superradiance to play an
important role in the scattering problem
for rapidly spinning black holes. For example, one could imagine 
that some partial waves which would otherwise 
be absorbed, could escape back to infinity. These waves might
then possibly make a noticeable contribution to the diffraction cross section,
provided that there were a sufficient number of them (as compared to the
total number $l_{\rm max}$ of partial waves contributing to the diffraction
scattering amplitude).

In order to investigate this possibility, 
we have performed a number of off-axis cross section calculations for a 
variety of wave frequencies $\omega M= 0.5 - 10$ and for $a \approx M$, 
i.e. black holes spinning near the extreme Kerr limit. 
We have found no qualitive difference whatsoever between those 
cross-sections and  the ones for a somewhat smaller spin value,  
$a=0.9M$ (say). In essence, we
were unable to find any effects in the cross section that could
be attributed to superradiance. Consequently, we are led to suspect 
that our intuition regarding the importance of superradiance for the 
scattering problem may be wrong.

This suspicion is confirmed by the following simple argument.
In order for a partial wave to be superradiant we should have 
$ 0 < \omega < m\omega_{+}$. Considering an extreme Kerr black hole (which
provides the best case for superradiant scattering) and 
the fact that $ m \leq l$,
we have the condition 
\begin{equation}
0 < 2\omega M < l
\label{sr1}
\end{equation}
As already mentioned, the partial waves for which superradiance 
will be important
are the ones with impact parameters $b \lesssim b_{\rm c}$, i.e. 
those that would be absorbed under different circumstances. 
This then requires that
\begin{equation}
l \lesssim \omega b_{\rm c} -1/2
\label{sr2}
\end{equation}
Combining (\ref{sr1}) and (\ref{sr2}) we arrive at the inequality
\begin{equation}
0 < 2M \lesssim b_{\rm c} - 1/2\omega
\label{sr3}
\end{equation}
The critical impact parameter (for prograde motion)  for an $a=M$ black
hole is $b_{\rm c} \approx 2M$. Hence the condition (\ref{sr3}) will not 
be satisfied, 
and it is unlikely that we would get a significant  number  of (if any) 
superradiant partial waves that could affect the cross section.

This conclusion may seem surprising given  results present 
in the literature \cite{book,Handler}. 
In particular, Handler and Matzner have briefly discussed 
the effect of superradiance on axially incident gravitational waves. 
They argue that
(see figure 14 in \cite{Handler}) ``superradiance has the effect of imposing 
a large background over the pattern, filling in the interference minima''.
Given our current level of understanding (or lack thereof) we cannot
at this point say whether superradiance can be the explanation for the 
effects observed by Handler and Matzner. After all, one should remember that
superradiant scattering strongly depends on the spin
of the field that is being scattered. It is well known \cite{Press}
 that gravitational perturbations can be amplified up to $138\%$ compared
to a tiny $0.04\%$ amplification for scalar fields (which is the case 
considered in this paper). This means that 
superradiance may significantly affect also partial waves with 
$b > b_{\rm c}$ in the gravitational wave case, which could
 lead to our simple argument not being valid.
This issue should be addressed by a detailed study of the 
scattering of gravitational waves from rotating black holes.
 

\section{Concluding Discussion}

We have presented an investigation of scattering of massless scalar waves
by a Kerr black hole. Our numerical work is based on phase-shifts
obtained via  integration of the relevant radial wavefunction with the help
of the Pr\"{u}fer phase-function method. This method has been shown
to be 
computationally efficient and to provide accurate results.
Using the obtained phase-shifts we have constructed differential
cross sections for several different cases. First we have discussed the
case of  waves incident along the black hole's rotation axis, for which we 
showed that
the resulting cross sections are similar to ones obtained
in the (non-rotating) Schwarzschild case. 
We then turned to the case of off-axis incidence, where the 
situation was shown to  change considerably. In that case the 
cross sections are generically asymmetric with
respect to the incidence direction. The overall diffraction pattern is
``frame dragged'', and as a result the backward glory maximum is shifted
along with of the black hole's rotation. Moreover, we have concluded that 
(at least for scalar waves) the so-called 
superradiance effect is unimportant for
 monochromatic scalar wave scattering. 

To summarize, our study provides a complete understanding of the 
purely rotational effects involved in black-hole scattering.   
Given this we are now well equipped to proceed to problems
of greater astrophysical interest, particularly ones
concerning gravitational waves. In these problems one would expect 
further features to arise as the spin and polarisation of the impinging
waves interact with the spin of the black hole. For incidence along 
the hole's spin axis, one can have circularly polarised waves 
which are either co- or counter-rotating. The two cases can lead to 
quite different results. Although the general features of 
the corresponding cross sections are similar, they show different
 structure in the backward direction \cite{book}. This is possibly due 
to interference between the two polarisation states of 
gravitational waves, an effect that has not yet been explored in detail.
Some initial work on gravitational-wave scattering has
been done, see \cite{Handler}, but we 
believe that the results of the present paper sheds new light 
on previous results, and could help interpret the 
rather complex cross sections that have been calculated in the 
gravitational-wave case.

In this context, it should be stressed that the choice of 
studying scalar waves was made solely on grounds of clarity and simplicity. 
Our approach can readily be extended to other cases.  
Moreover, it is relevant to point out that
a full off-axis gravitational wave scattering cross section 
calculation is still missing. We would expect such cross 
sections to be rather complicated, combining  the 
frame-dragging effects discussed in this
paper with various spin-induced features. We hope to be able to study this
interesting problem in the near future. 

\acknowledgements

K.G. thanks the State Scholarships Foundation of Greece for financial support.
N.A. is a Philip Leverhulme Prize Fellow, and also
acknowledges support from PPARC via grant number PPA/G/1998/00606 and 
the European Union via the network ``Sources for Gravitational Waves''.

\appendix


\section{Plane waves in the Kerr geometry}

The long-range character of the gravitational field
modifies the form of  ``plane waves''. This non-trivial
issue has been discussed in the context of black hole scattering by
Matzner \cite{Matzner} and Chrzanowski {\it et al.} \cite{Chrzanowski}. 
For completeness, we provide a brief discussion here.
   
In a field-free region a monochromatic plane wave is, of course, given by 
the familiar expression
\begin{equation}
\Phi_{\rm plane}= e^{i\omega r \cos\theta -i\omega t}
\label{fplane}
\end{equation}
when a spherical coordinate frame is employed. The plane wave is taken
to travel along the $z$-axis.
The field in (\ref{fplane}) solves the the wave equation 
$\Box\Phi_{\rm plane}=0 $. Moreover, we can assume a decomposition of the form
\begin{equation}
\Phi_{\rm plane}= \frac{e^{-i\omega t}}{r} \sum_{l} c_{l}^{(0)} u_{l}^{(0)}
(r) P_{l}(\cos\theta) e^{-i\omega t}
\end{equation}
where the radial wavefunction satisfies
\begin{equation}
\frac{d^2 u_{l}^{(0)}}{dr^2} + \left [ \omega^2 -\frac{l(l+1)}{r^2} 
\right ] u_{l}^{(0)}= 0
\end{equation}
Let us now consider a ``plane wave'' in the Schwarzschild geometry.
First of all, we  expect  such a field to be only an asymptotic solution
(as $r \to \infty $)  of the full wave equation $\Box \Phi=0$
(where  $\Box$ represents the covariant d'Alembert operator).        
The plane wave field can then be represented at infinity as
\begin{equation}
\Phi_{\rm plane} \approx \frac{e^{-i\omega t}}{r} \sum_{l} c_{l}^{(0)} u_{l}^{(0)}
(r) P_{l}
(\cos\theta) e^{-i\omega t}
\end{equation}
and
the radial wavefunction  will be a solution of
\begin{equation}
\frac{d^2 u_{l}^{(0)}}{dr_{\ast}^2} + \left [ \omega^2 -
\frac{l(l+1)}{r_{\ast}^2} + {\cal O} \left ( \frac{\ln r_{\ast}}{r_{\ast}^3}
\right ) \right ] u_{l}^{(0)}= 0
\end{equation}
This equation is similar to the corresponding flat space equation.
The only 
 difference is the appearance of the tortoise coordinate $r_{\ast}$
instead of $r$. Hence, we are inspired to write the plane wave field as
\begin{equation}
\Phi_{\rm plane}= e^{i\omega r_{\ast} \cos\theta -i\omega t}
\label{splane}
\end{equation}
From this discussion, it
should be clear that this form is valid only for $r \to \infty$.
It is straightforward to see that in the same regime (\ref{splane}) 
solves $\Box\Phi=0$. 
Expression (\ref{splane}) is the closest we can get to the usual plane wave
form (\ref{fplane}). The long-range gravitational field is simply taken
into account by an appropriate phase modification.

Next, we consider a plane wave in Kerr geometry. For simplicity we take the
$z$-axis to coincide with the black hole's spin axis. We can then write
\begin{equation}
\Phi_{\rm plane} \approx \frac{e^{-i\omega t}}{r} \sum_{l} c_{l}^{(0)}
u_{l}^{(0)}(r) S_{l0}^{a\omega}(\theta) e^{im\varphi}
\label{kplane}
\end{equation}
The radial wavefunction is  solution of
\begin{equation}
\frac{d^2 u_{l}^{(0)}}{dr_{\ast}^2} + \left [ \omega^2  -
\frac{\lambda + 2am\omega}{r_{\ast}^2} + {\cal O} \left (\frac{\ln r_{\ast}}
{r_{\ast}^3} \right ) \right ] u_{l}^{(0)}=0
\label{kplane2}
\end{equation}
It is obvious that both (\ref{kplane}) and (\ref{kplane2}) are  different
from the corresponding flat space expressions. That is, unlike in the 
Schwarzschild case, we are not able to derive an explicit form for a plane
wave. Thus, we \underline{postulate} the following asymptotic expression for a 
plane wave travelling along the $z$-axis
\begin{equation}
\Phi_{\rm plane}= e^{i\omega r_{\ast}\cos\theta -i\omega t}
\label{kplane3}
\end{equation}
where $r_{\ast}$ is the appropriate tortoise coordinate (\ref{rstar}). 
The field given by (\ref{kplane3}) is a solution of $\Box \Phi=0$
for $r \to \infty$.
For the general case of a plane wave travelling along a direction that 
makes an angle $\gamma$ with the $z$-axis the appropriate 
expression is given by
(\ref{ofplane}).

 
\section{Asymptotic expansion of plane waves}

In this Appendix the asymptotic expansion of a plane wave in Kerr background
is worked out. The calculation presented here is identical to the one found
in the Appendix A1 of \cite{book}, but here it is specialised to $s=0$. 
We have seen that for $r \to \infty $ the plane wave decomposition
becomes
\begin{equation}
e^{i\omega r_{\ast} [\sin\gamma\sin\theta\sin\varphi +\cos\gamma\cos\theta]}
\approx \frac{1}{\omega r} \sum_{l,m} c_{lm}^{(0)} u_{lm}^{(0)}(r \to \infty)
S_{lm}^{a\omega}(\theta) e^{im\varphi}
\end{equation}
where $\gamma$ is the angle between the wave's propagation direction 
and the positive $z$-axis. We multiply this expression by 
$S_{l^{\prime}m^{\prime}}^{a\omega}(\theta^{\prime})e^{-im^{\prime}\varphi^
{\prime}}$ and integrate over the angles to get (after a trivial change
$l^{\prime} \to l$, $ m^{\prime} \to m$ at the end)
\begin{equation}
c_{lm}^{(0)} u_{lm}^{(0)}(r \to \infty) \approx \omega r \int_{0}^{\pi}
d\theta \sin\theta ~ S_{lm}^{a\omega}(\theta) e^{i\omega r_{\ast} \cos\gamma
\cos\theta } \int_{0}^{2\pi} d\varphi e^{i\omega r_{\ast} \sin\gamma \sin\theta
\sin\varphi -im\varphi }
\label{c1}
\end{equation}
The integration over $\varphi$ can be performed with  a little help
from  \cite{Gradshteyn}, and the  result is
\begin{equation}
\int_{0}^{2\pi} d\varphi e^{i\omega r_{\ast} \sin\gamma \sin\theta
\sin\varphi -im\varphi }= 2\pi J_{m}(\omega r_{\ast} \sin\gamma \sin\theta )
\end{equation}
Since the Bessel function has a large argument it can be approximated as
\cite{Stegun},
\begin{equation}
J_{\nu}(z) \approx \frac{1}{\sqrt{2\pi z}} \left ( e^{i(z- \nu\pi/2 -\pi/4)}
+ e^{-i(z -\nu\pi/2 -\pi/4)} \right )
\end{equation}
Note that this approximation is legal as long as $\gamma \neq 0 $. The 
on-axis case $\gamma=0$ can be treated seperately, in a way similar to the
one sketched here.  
Using this approximation in (\ref{c1}) we get
\begin{eqnarray}
& & c_{lm}^{(0)}u_{lm}^{(0)}(r \to \infty) \approx  \sqrt{\frac{2\pi\omega r}
{\sin\gamma} } \left ( e^{-\frac{i}{2}(m\pi + \pi/2)} {\cal I}_{-}
+  e^{\frac{i}{2}(m\pi + \pi/2)} {\cal I}_{+} \right ) \\
&& {\cal I}_{\pm} = \int_{0}^{\pi} d\theta \sqrt{\sin\theta} S_{lm}^{a\omega}
(\theta) e^{i\omega r_{\ast} \cos(\theta \pm \gamma) }
\end{eqnarray}
The ${\cal I}_{\pm} $ integrals can be evaluated using the stationary phase
approximation. We obtain
\begin{equation}
c_{lm}^{(0)} u_{lm}^{(0)} (r \to \infty) \approx 2\pi  \left[ (-i)^{m+1}
e^{i\omega r_{\ast}} S_{lm}^{a\omega}(\gamma) +  i^{m+1} e^{-i\omega r_{\ast} } S_{lm}^{a\omega}(\pi -\gamma)  \right]
\end{equation}
We  finally get (\ref{plane_asym}) by using the symmetry relation
\begin{equation}
S_{lm}^{a\omega}(\pi -\theta)= (-1)^{l+m} S_{lm}^{a\omega}(\theta)
\end{equation}


\section{Calculation of spheroidal harmonics and their eigenvalues}

For the numerical calculation of the spheroidal harmonics we have adopted
a ``spectral decomposition'' method, first developed by Hughes \cite{Scott} in the context of gravitational wave emission and radiation
 backreaction on particles orbiting rotating black holes. 
In the present work we have
specialised this technique for the spin-$0$ spheroidal harmonics. 
The angular equation satisfied by $S^{a\omega}_{lm}(\theta)$ is,
\begin{equation}
\frac{1}{\sin\theta} \frac{d}{d\theta} \left ( \sin\theta \frac{d S_{lm}^
{a\omega}}{d\theta} \right ) + [ (a\omega)^2 \cos^{2}\theta -\frac{m^2}
{\sin^{2}\theta} + E_{lm} ]S^{a\omega}_{lm}=0
\label{spheroid}
\end{equation}   
where $E_{lm}$ denotes the corresponding eigenvalue. For the special case
$a\omega=0$ we have $E_{lm}=l(l+1)$ and the solution of (\ref{spheroid})
is the familiar spherical harmonic (here we are suppressing the 
dependence on $\varphi$)
\begin{equation}
Y_{lm}(\theta)= 
\left [ \frac{2l+1}{4\pi} \frac{(l-m)!}{(l+m)!} \right ]^{1/2}
P_{lm}(\cos\theta)
\end{equation}
where $P_{lm}$ is the associated Legendre polynomial. It's numerical
calculation is quite straigthforward \cite{Recipes} based on the recurrence
relation
\begin{equation}
P_{lm}(x)=  \frac{1}{l-m} [x(2l-1)P_{l-1,m} -(l+m-1)P_{l-2,m} ]
\end{equation} 
with ``initial conditions''
\begin{eqnarray}
P_{mm}(x)&=&(-1)^m (2m-1)!!(1-x^2)^{m/2}    \\
P_{m+1,m}(x) &=& (2m+1)xP_{mm}(x) 
\end{eqnarray}
We can always expand the spheroidal harmonic in terms of spherical harmonics,
\begin{equation}
S_{lm}^{a\omega}(\theta)= \sum_{j=|m|}^{\infty} b_{j}^{a\omega} Y_{jm}(\theta)
\label{sp_dec}
\end{equation}
Substituting this spectral decomposition in (\ref{spheroid}), multiplying
with $Y_{lm}(\theta) $ and integrating over $\theta$ we get
\begin{equation}
(a\omega)^2 \sum_{j=|m|}^{\infty} b_{j}^{a\omega} 
c_{jl2}^{m} -b_{l}^{a\omega} l(l+1) = -E_{lm} 
b_{l}^{a\omega} 
\label{step1}
\end{equation} 
where 
\begin{equation}
c_{jl}^{m}= 2\pi \int_{0}^{\pi} d\theta \sin\theta \cos^2\theta ~
Y_{lm}(\theta)Y_{jm}(\theta)
\end{equation}
This integral can be evaluated in terms of Clebsch-Gordan coefficients
\cite{Stegun}
\begin{equation}
c_{jl}^{m}= \frac{1}{3} \delta_{lj} + \frac{2}{3} \sqrt{\frac{2j+1}{2l+1}}
<j2m0|lm> <j200|l0>
\end{equation}
It follows that $c_{jl}^{m} \neq 0$ only for $ j= l-1,l,l+1$. Then (\ref{step1})
gives
\begin{equation}
[(a\omega)^2 c_{l-2,l}^{m} ]b_{l-2}^{a\omega} + [(a\omega)^2 c_{l,l}^{m}
-l(l+1) ] b_{l}^{a\omega} + [(a\omega)^2 c_{l+2,l}^{m} ] b_{l+2}^{a\omega}=
-E_{lm} b_{l}^{a\omega}
\label{step2}
\end{equation}
We can rewrite (\ref{step2}) as an eigenvalue problem for the matrix
$M_{ij}= (a\omega)^2 c_{ji}^{m}$ with eigenvector $b^{i}= b^{a\omega}_{i}$
and eigenvalue $E_{lm}$. Clearly, ${\bf M}$ is a real band-diagonal matrix.
Standard routines from \cite{Recipes} can be employed to find the
eigenvectors and eigenvalues of such a matrix. Then, the spheroidal harmonic 
is directly obtained from (\ref{sp_dec}) (even though it involves an infinite sum, in reality only few coefficients $b_{j}^{a\omega}$ are significant).
The described spectral decomposition method is reliable, unless $a\omega$
becomes large compared to unity (under such conditions the matrix ${\bf M}$
is no longer diagonally dominant, and the convergence of the method is very 
slow). In effect, very high frequency cross sections for Kerr scattering
will be inaccurate (especially when the black hole is rapidly rotating).


\begin{thebibliography}{}


\bibitem{book}
        J.A.H. Futterman, F.A. Handler and R.A. Matzner,
\textit{Scattering from Black Holes} (Cambridge University Press,
Cambridge, England, 1988).

\bibitem{Ryan} 
       R.A. Matzner and M.P. Ryan, Jr., Astrophys. J. Suppl. {\bf 36},
      451 (1978).

\bibitem{Sanchez1}
        N. Sanchez, J. Math. Phys. {\bf 17}, 688 (1976)


\bibitem{Sanchez2}
        N. Sanchez, Phys. Rev. D  {\bf 16}, 937 (1977)


\bibitem{Sanchez3}
        N. Sanchez, Phys. Rev. D {\bf 18}, 1030 (1978)

\bibitem{Sanchez4}
        N. Sanchez, Phys. Rev. D {\bf 18}, 1798 (1978)

\bibitem{path_int1}
        P. Anninos, C. DeWitt-Morette, R.A. Matzner, P. Yioutas, and
        T-R Zhang, Phys. Rev. D {\bf 46}, 4477 (1992)

\bibitem{Nils} 
        N. Andersson, Phys. Rev. D {\bf 52}, 1808 (1995)


\bibitem{Handler}
        F.A. Handler and R.A. Matzner, Phys. Rev. D {\bf 22}, 2331 (1980)


\bibitem{Newton}
        R.G. Newton, \textit{Scattering Theory of Waves and Particles}
        (McGraw-Hill, New York, 1966).


\bibitem{CAM1}
        N. Andersson and K-E. Thylwe, Class. Quantum Grav. {\bf 11},
       2991 (1994)

\bibitem{CAM2} 
        N.Andersson, Class. Quantum Grav. {\bf 11}, 3003 (1994)
        

\bibitem{path_int2}
         C. DeWitt-Morette and B.L. Nelson, Phys. Rev. D {\bf 29}, 1663 (1984)

\bibitem{path_int3} T-R Zhang and C. DeWitt-Morette, Phys. Rev. Lett. {\bf 52},
         2313 (1984)
        

\bibitem{Froman1} 
       N. Fr\"{o}man, P.O. Fr\"{o}man, and B. Lundborg, Math. Proc.
       Cambridge Philos. Soc. {\bf 104}, 153 (1988)

\bibitem{Froman2} 
        N. Fr\"{o}man, P.O. Fr\"{o}man, in \textit{Forty More Years of
        Ramifications: Spectral Asymptotics and its Applications},
    edited by S.A. Fulling and F.J.Narcowich, Discourses in Mathematics
and its Applications, No. 1(Texas A\& M University, Department of Mathematics,
1992), pp 121-159

\bibitem{Pajunen1}
         P. Pajunen, J. Chem. Phys. {\bf 88}, 4268, (1988)

\bibitem{Pajunen2}
         P. Pajunen, J. Comp. Phys. {\bf 82}, 16, (1989)

\bibitem{Pryce}
         J.D. Pryce, \textit{Numerical Solution of Sturm-Liouville Problems}
         (Oxford Science Publications, Oxford, England, 1993)  

\bibitem{pam}
	N. Andersson, Proc. R. Soc. Lon. A {\bf 439}, 47, (1992)

\bibitem{Teukolsky}
      S.A. Teukolsky, Astrophys. J. {\bf 185}, 635 (1973)


\bibitem{Press}
      S.A. Teukolsky and W.H. Press, Astrophys. J. {\bf 193}, 443 (1974)

\bibitem{Matzner}
      R.A. Matzner, J.Math.Phys. {\bf 9}, 163 (1968)


\bibitem{Chrzanowski}
      P.L. Chrzanowski, R.A. Matzner, V.D. Sandberg, and M.P. Ryan,
      Jr. , Phys. Rev. D {\bf 14}, 317 (1976) 


\bibitem{Ford}
         K.W. Ford and J.A. Wheeler, Ann. Phys. (N.Y.) {\bf 7}, 287 (1959)


\bibitem{Chandra}
       S. Chandrasekhar, \textit{The Mathematical Theory of Black Holes}
(Oxford University Press, New York, 1983)


\bibitem{Recipes}
        W.H. Press, S.A. Teukolsky, W.T. Vetterling and B.P. Flannery, 
\textit{Numerical Recipes} (Cambridge University Press, Cambridge, England,
1992).

\bibitem{Stegun}
        M. Abramowitz and I.A. Stegun \textit{Handbook of Mathematical 
       Functions} (Dover Publications INC., New York, 1985)

\bibitem{Gradshteyn}
       I.S. Gradshteyn and I.M. Ryzhik, \textit{Tables of Integrals, Series
and products} (Academic Press INC., London 1980)

\bibitem{Scott}
      S.A. Hughes, Phys. Rev. D {\bf 61}, 084004 (2000)

 


\end{thebibliography}
\end{document}